\documentclass[onecolumn]{aastex631}
\usepackage{xpatch}
\xpatchcmd{\thebibliography}{\twocolumngrid}{}{}{}
\usepackage{graphicx}
\usepackage{psfrag}
\usepackage{amsmath}
\usepackage{amssymb}
\usepackage{color,xcolor}
\usepackage{hyperref}
\hypersetup{
    colorlinks=true,
    citecolor=blue,
    filecolor=black,
    linkcolor=blue,
    urlcolor=magenta,
    linktocpage=true,
    breaklinks = true
}

\usepackage{enumitem}
\usepackage{subfigure,float}
\usepackage{multirow}
\usepackage{makecell}
\usepackage[normalem]{ulem}

\usepackage{array,booktabs}
\usepackage{mathtools}
\DeclarePairedDelimiter\abs{\lvert}{\rvert}

\def\apjl{Astrophys. J.}

\def\aap{Astron. Astrophys.}
\def\apjs{Astrophys. J. Suppl.}

\newcommand{\beq}{\begin{equation}}
\newcommand{\eeq}{\end{equation}}

\newcommand{\ber}{\begin{eqnarray}}
\newcommand{\eer}{\end{eqnarray}}

\newcommand{\ba}{\begin{align}}
\newcommand{\ea}{\end{align}}

\def \dte {\Delta t_{\rm est}}

\def \mut {\widehat{\mu}}
\def \dtt {\widehat{\Delta t}}

\def \muc {\mu}
\def \dtc {\Delta t}

\def \f {f}
\def \ept {\epsilon(\dtc)}

\def \dt{\Delta t}

\def \F{\mathcal{F}}

\def \p {\color{purple}}

\def \A {\mathcal{A}}

\def \frI {f_{\rm rec}}

\def \tc {t_c}
\def \acf {{\rm ACF}}
\def \acff {{\rm ACF}(F;\dtc)}
\def \acfh {{\rm ACF}(H;\dtc)}
\def \acffi {{\rm ACF}(f;\dtc)}
\def \acfhi {{\rm ACF}(h;\dtc)}
\def \acfX {{\rm ACF}(X;\dtc)}
\def \acfx {{\rm ACF}(x;\dtc)}

\def \epc {\epsilon(\dtc)}
\def \Se {\Sigma_{\epsilon(\dtc)}}
\def \SH {\Sigma_{\acfh}}

\def \htI {h}
\def \tr {t_{\rm range}}
\def \epo {\epsilon_0}

\begin{document}
\title{Harnessing Unresolved Lensed Quasars: The Mathematical Foundation of the Fluctuation Curve}

\author[0000-0003-0141-606X]{Satadru Bag}
\email{satadru@kasi.re.kr}
\affiliation{Korea Astronomy and Space Science Institute, Daejeon 34055, Korea}

\author[0000-0002-6039-8247]{Wuhyun Sohn}
\affiliation{Korea Astronomy and Space Science Institute, Daejeon 34055, Korea}

\author[0000-0001-6815-0337]{Arman Shafieloo}
\email{shafieloo@kasi.re.kr}
\affiliation{Korea Astronomy and Space Science Institute, Daejeon 34055, Korea}
\affiliation{University of Science and Technology, Daejeon 34113, Korea}

\author[0000-0002-4359-5994]{Kai Liao}
\affiliation{School of Physics and Technology, Wuhan University, Wuhan 430072, China}

\begin{abstract}
Strong gravitational lensed quasars (QSOs) have emerged as powerful and novel cosmic probes as they can deliver crucial cosmological information, such as a measurement of the Hubble constant, independent of other probes. Although the upcoming LSST survey is expected to discover $10^3-10^4$ lensed QSOs, a large fraction will remain unresolved due to seeing. The stochastic nature of the quasar intrinsic flux makes it challenging to identify lensed ones and measure the time delays using unresolved light curve data only. In this regard, \citet{Bag2021qso} introduced a data-driven technique based on the minimization of the fluctuation in the reconstructed image light curves.  In this article, we delve deeper into the mathematical foundation of this approach. We show that the lensing signal in the fluctuation curve is dominated by the auto-correlation function (ACF) of the derivative of the joint light curve. This explains why the fluctuation curve enables the detection of the lensed QSOs only using the joint light curve, without making assumptions about QSO flux variability, nor requiring any additional information. We show that the ACF of the derivative of the joint light curve is more reliable than the ACF of the joint light curve itself because intrinsic quasar flux variability shows significant auto-correlation up to a few hundred days (as they follow a red power spectrum). In addition, we show that the minimization of fluctuation approach provides even better precision and recall as compared to the ACF of the derivative of the joint light curve when the data have significant observational noise.
\end{abstract}

\section{Introduction}

Strong gravitational lensed systems have emerged as a powerful and novel cosmic probe (see, e.g.,\citet{2016A&ARv..24...11T} for a review). They can deliver cosmological information independent of other probes such as the Type Ia supernovae (SNe), Baryon Acoustic Osculations (BAO), and Cosmic Microwave Background (CMB). Time delay measurements, together with accurate lens modelling, allow us to directly estimate the present epoch value of the cosmic expansion rate, i.e. the Hubble constant ($H_0$) \citep{Refsdal1964_1,Refsdal1964_2,Saha,Oguri_2007, Bonvin2017,Wong:2019kwg,2020A&A...643A.165B,Birrer}. Therefore, `time delay cosmography' can play a crucial role in elucidating the ongoing $H_0$ tension between the local measurements \citep{Riess2022,Abdalla:2022} and  early universe probes like the CMB \citep{Planck:2018vyg}. Other applications of strong lensing in cosmology and astrophysics are summarised in the review by~\citet{Treu2010}.
 
For time delay measurements, one needs time-variable sources, such as quasars (QSOs) and supernovae (SNe). Lensed SNe\citep{2019RPPh...82l6901O,2022ChPhL..39k9801L,2023arXiv230107729S} are extremely rare as only four with multiple images have been discovered so far \citep{Kelly:2014mwa,Goobar:2016uuf,2021NatAs.tmp..164R, Sn_zwicky}. In comparison, lensed QSOs are more abundant and thus remain to be the primary source for the time delay cosmography (however, lensed SNe could be at the forefront of time delay cosmography in the next decade \citep{2020A&A...644A.162S}). Although hundreds of lensed QSOs are known \citep{Lemon2022}, only a few have been used for cosmology. For example, using only six `good quality' lensed QSOs the H0LiCOW team \citep{Suyu:2016qxx} measured the Hubble constant with $2.4\%$ uncertainty~\citep{2020MNRAS.498.1420W}, under standard assumptions about the mass density profile of the deflector, and a seventh brings the precision to 2\% \citep{Shajib2020,Millon2020}. However, if one drops the assumptions and adopts density profiles that are maximally degenerate with $H_0$ through the mass sheet degeneracy, the uncertainty increases to $9\%$ \citep{2020A&A...643A.165B}, highlighting the need for substantially larger samples.  While the uncertainty can also be reduced by additional information per lens (especially stellar kinematics), a powerful way of improving the $H_0$ precision is to increase the sample volume significantly~\citep{2021A&A...656A.153S}. For example, observations of hundreds of  lensed systems will deliver sub-percent uncertainty and accuracy, regardless of any assumption on their mass density profile \citep{Birrer,2016JCAP...04..031J}. 
 
The angular separation of images are typically of the order $\sim 1-2$ arcsec for  galaxy scale lenses \citep{1996astro.ph..6001N,Treu2010}. Therefore, one is required to first resolve the images by using either a sufficiently high-resolution (ground-based or space) telescope or through spectroscopy. Then the individual light curves need to be monitored at sufficient resolution for several years in order to measure the time delays \citep[e.g.,][]{Tewes2013,2015ApJ...800...11L,Millon:2020xab}. This can be 
difficult as these observations are expensive. On the other hand, unresolved light curves may be available ``for free" from synoptic surveys. For example, we expect a lot of lensed QSOs to be partially resolved or completely unresolved in the wide field surveys, such as Zwicky Transient Facility (ZTF) \citep{ztf} and Legacy Survey of Space and Time (LSST) \citep{lsst1,lsst2}, because their angular resolution is limited by seeing.
In such cases, we can observe the joint light curve of the unresolved system that is a blend of the individual light curves. A robust method of detecting lensed QSOs through unresolved light curves can take advantage of the more abundant smaller telescopes. Thus, the importance of this approach cannot be overstated for boosting the sample size of the observed lensed systems \footnote{Although this work focuses on the unresolved lensed QSOs, similar exercises for the unresolved lensed supernovae have been pursued in the literature \citep{Bag:2020pbg,Denissenya:2021cpz,2022MNRAS.515..977D}.}. 
 
 There are multiple other advantages in working with unresolved systems. Since there is no need for resolving the images a priori, this approach can be applied to the light curve data from ongoing time domain surveys such as ZTF \citep{ztf}, Pan-STARRS1 \citep{Pan-STARRS1}. This will become more crucial when the upcoming Vera Rubin Observatory starts 
 the LSST \citep{lsst1,lsst2}. This approach also evades any degeneracy between a binary QSO pair and a doubly lensed QSO that creates confusion in lens detection using the resolved photometry \citep{Peng_1999,Mortlock1999}.

Recently, multiple different techniques have been proposed to identify the lensed systems and to measure their time delays using the joint light curves \citep{Shu2020,Springer:2021jhm,Springer:2021yhe,2022MNRAS.515.5665B}.
The primary challenge in detecting the lensed cases using the joint light curves is that the intrinsic QSO light curves are highly stochastic and show a vast diversity in the flux variability. Therefore, any assumption on the flux variability can lead to biased results with low precision and high false positive detection rate when the real light curves are not well described by the assumption. Therefore, it is extremely important to be model agnostic for achieving higher recall and precision as well as for reliable unbiased results.

In paper 1 \citep{Bag2021qso} we introduced a novel data-driven method that can detect the lensed QSOs and simultaneously measures the time delays only using the joint light curves, most importantly neither assuming anything about the quasar flux variability nor using any additional information. The technique is based on the empirical observation that the reconstructed image light curves corresponding to incorrect time delays exhibit more fluctuation than the ones reconstructed using the correct time delay. Although \citet{Bag2021qso} demonstrates that this approach is successful in the presence of significant noise (e.g. ZTF-like noise) and on existing data quality, it lacks a formal explanation as to how the minimization of the fluctuation works.
This article looks deeper into the mathematical formalism of this approach. We attempt to understand the mathematical reasoning behind the empirical observations which laid the foundation of this approach. A clear insight into the mechanism should allow us to explore the strengths and possible limitations of this approach. 

The paper is organized as follows. Section \ref{sec:method} recapitulates the method introduced by \citet{Bag2021qso}. In Section \ref{sec:math_proof}, we provide a detailed mathematical explanation for the method's ability to detect lensed systems. We show that the lensing signal in the fluctuation curve is dominated by the auto-correlation function (ACF) of the derivative of the joint light curve. Section \ref{sec:acff_acfh} explains why the ACF of the derivative of the joint light curve performs better in identifying lensed systems than the ACF of the joint light curve itself. Finally, in Section \ref{sec:fluc_vs_acfh} we compare the performance of ACF of the derivative of the joint light curves against the full fluctuation curves. We conclude our findings in the Section \ref{sec:conclusion}.

\section{Reconstructing the underlying light curves and fluctuations in them}
\label{sec:method}
The joint light curve of an unresolved lensed QSO having $N_I$ images is the sum of the image light curves,
\beq \label{eq:gen_lensed_flux}
F(t)=\sum_{j=1}^{N_I} a_j \mathcal{F}(t-T_j)\;,
\eeq
where the individual image light curves can be described by a common function $\mathcal{F}(t)$ but with different magnifications ($a_j$) and time delays ($T_j$). To break the degeneracy among the images, we choose the ordering such that $a_1 \ge a_2 \ge \cdots \ge a_{N_I}$ without loss of generality. 
For simplicity, let us first consider double systems (two images) with
\beq \label{eq:lensed_flux}
F(t)=\f(t)+\mut \f(t-\dtt)~, {\rm where}~~\f(t)=a_1\mathcal{F}(t-T_1)~,~\mut=a_2/a_1~{\rm and}~\dtt=T_2-T_1\;.
\eeq
Here, $\f(t)$ is the light curve of the brighter image which we take as a reference. The other (fainter) image's light curve is then given by $\mut f(t-\dtt)$, where $\mut \le 1$ and $\dtt$ correspond to the magnification ratio and time delay, respectively, with respect to the reference (brighter) image. Note that $\dtt$ can be positive or negative; a positive (negative) $\dtt$ implies that the fainter image arrives later (earlier) in time than the brighter image.

It is difficult to model the quasar flux variability due to its highly stochastic nature. Instead, one can reconstruct the light curve of the brighter image 
following \citet{Bag2021qso} \cite[see also][]{Geiger1996} as
\begin{align}
\frI(t;\muc,\dtc)&=F(t)-\mu \frI(t-\dt;\muc,\dtc) \;, \label{eq:eqF1} \\ 
&=F(t)-\mu F(t-\dt)+\mu^2 F(t-2\dt) \nonumber \\ 
& \qquad \qquad -\mu^3F(t-3\dt)+\mu^4F(t-4\dt)- \hdots\;, \nonumber \\ 
      &= \sum_{n=0}^{\infty} \left( -\mu \right)^n F(t-n\cdot \dt)\;,\label{eq:rec}   
\end{align}
from the joint light curve $F(t)$, given $\dtc$ and $\muc<1$ which need not be equal to the true underlying values. In this article we denote the true magnification ratio and time delay by $\lbrace \mut,~\dtt \rbrace$ to avoid confusion with the free parameters of the reconstruction, $\lbrace \muc,~\dtc \rbrace$, which are same as $\lbrace \mu_{\rm try},~\dt_{\rm try} \rbrace$ in the paper 1 \citep{Bag2021qso}.

Note that the higher order terms in Eq. \eqref{eq:rec} require $F(t)$ outside its observed range. Since the quasar flux cannot be predicted, we assume that $F(t)$ remains flat outside its observed range. As discussed in \citet{Geiger1996, Bag2021qso}, this convention has a negligible effect on the reconstruction except near either of the boundaries (depending upon the sign of $\dtc$).

We emphasise that by restricting ourselves to $\muc<1$, that ensures the convergence of the sum in Eq. \eqref{eq:rec}, we are essentially reconstructing the light curve of the brighter image without any loss of generality.
Moreover, for any choice of $\lbrace \muc<1,~\dtc \rbrace$, Eq. \eqref{eq:rec} gives a unique solution for the brightest image light curve; the corresponding fainter image light curve would be $\muc \frI(t-\dtc;\muc,\dtc)$. Only when $\muc=\mut$ and $\dtc=\dtt$, however, Eq. \eqref{eq:rec} recovers the true image light curve, $\frI(t)=\f(t)$. Still, the equation $F(t)=\frI(t;\muc,\dtc)+\muc \frI(t-\dtc;\muc,\dtc)$ is exactly satisfied for any choice of $\lbrace \muc<1,~\dtc \rbrace$ by construction. This demonstrates the mathematical degeneracy present in this lensed detection problem as discussed in \citet{Geiger1996,Bag2021qso}; without any prior assumptions on $\f(t)$, any choice of $\lbrace \muc,~\dtc \rbrace$ can yield a (unique) lensing solution to the joint light curve $F(t)$.

 In paper 1 \citep{Bag2021qso}, we introduce a completely data-driven technique to break the degeneracy in the time delay by minimizing the fluctuation in reconstructed image light curves. To quantify the amount of fluctuation in a reconstructed image light curve  $\frI(t;\muc,\dtc)$, \citet{Bag2021qso} uses the simple metric,
\beq\label{eq:ep_def}
 \epsilon (\dtc;\muc)\equiv\sum_{i=0}^{N_D -1}\left[\frI(t_{i+1};\muc,\dtc)-\frI(t_{i};\muc,\dtc) \right]^2\;.
 \eeq
Fixing the trial magnification ratio ($\muc$) to an arbitrary value (less than unity), one reconstructs the brighter image light curve for a number of $\dtc$ and finally looks for minima in the fluctuation curve, $\epsilon(\dtc)$.

 \begin{figure}
 \centering
\includegraphics[width=\linewidth]{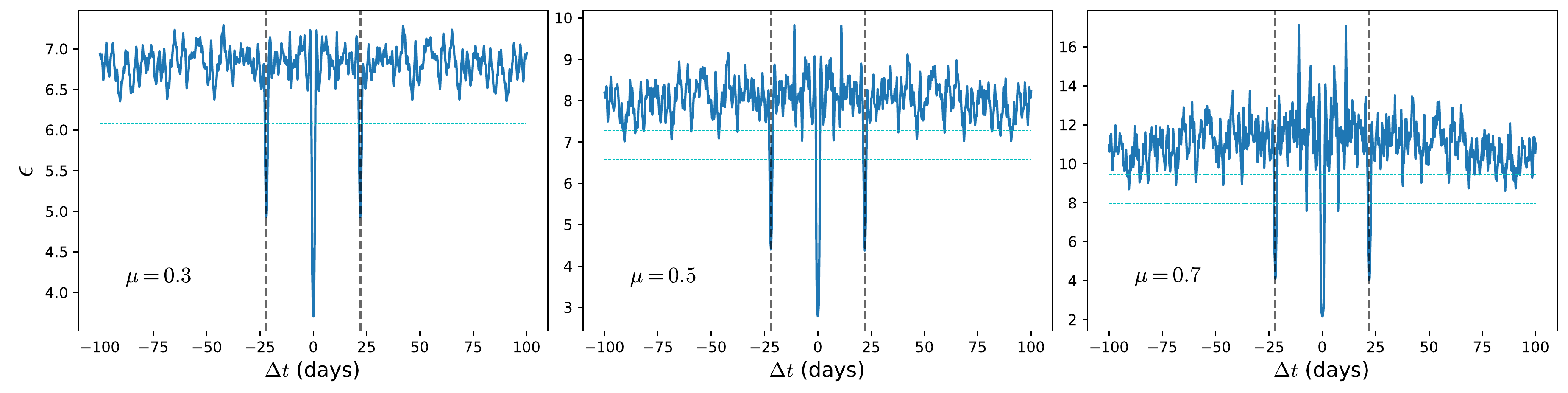}
\caption{The fluctuation curves for an example of a double system simulated (in perfect condition with negligible observation noise) using damped random walk (DRW) process with time delay $\dtt=22.0$ days and the magnification ratio $\mut=0.86$. The amount of fluctuation ($\epsilon$ calculated from Eq. \eqref{eq:ep_def}) in the reconstructed image light curves is plotted against the trial $\dtc$. The three panels correspond to three arbitrary choices of trial magnification ratio, $\muc=0.3,0.5,0.7$. The dashed vertical lines in each plot mark the true time delay, $\dtc=\pm \dtt$, where we find prominent pairs of secondary minima that can be used to detect the lensed system and measure the time delay. This figure is qualitatively similar to Fig. 2 of \citet{Bag2021qso}.}
\label{fig:ep_demo}
\end{figure}

Fig.~\ref{fig:ep_demo} demonstrates how one can detect the unresolved lensed QSOs through the fluctuation curves by considering an example of a double (2-image) system that is simulated using the {\em damped random walk} (DRW) process in perfect conditions (marginal observational noise and 1 day of cadence). The true magnification ratio and time delay are set to: $\mut=0.86$ and $\dtt=22.0$ day.  The three panels show $\ept$ for different (arbitrary) choices of $\muc$.  We make the following observations from Fig.~\ref{fig:ep_demo}.
\begin{itemize}
 \item The fluctuation curve $\epsilon(\dtc)$ is highly symmetric with respect to $\dtc=0$.
 \item There exists a global minimum at $\dtc=0$;  the amount of fluctuations in the reconstructed light curve is minimised when we assume that the system is not lensed. This is a generic feature of all fluctuation curves irrespective of the system being lensed or not, and hence can be ignored \footnote{While trying to detect the lensed QSOs through the minimal fluctuation, as proposed by \citet{Bag2021qso}, this unlensed solution always remains to be a feasible lensing solution to the unresolved problem in the absence of any assumption on the flux variability.}. 
 \item Strikingly, when $\dtc=\pm\dtt$ (shown by the dashed vertical lines in each panel), we observe a pair of prominent secondary minima that correctly identify the system as lensed and simultaneously estimate the time delay accurately. On the other hand, if the system is unlensed (i.e. $\mut =0$ or $\dtt=0$) we do not find any prominent secondary minimum in $\epsilon(\dtc)$ as demonstrated in \citet{Bag2021qso}.
 \item All of the observations above are somewhat insensitive to the choice of $\muc$. Therefore, the true magnification ratio cannot be estimated in this approach. Nevertheless, one can accurately measure the time delays from the location of the pair of secondary minima in the fluctuation curves.
\end{itemize}
In the next section, we provide the mathematical reasoning behind all the above empirical observations. 
 
 \section{Why is fluctuation minimised for correct time delay?}
 \label{sec:math_proof}
Substituting Eq. \eqref{eq:rec} into Eq. \eqref{eq:ep_def}, one gets 
 \begin{align}
 \epsilon (\dtc) &= \sum_{i}^{N_D -1}\left[ \sum_n^{\infty} \lbrace \left(-\muc \right)^n F(t_{i+1}-n\dtc) \rbrace - \sum_m^{\infty} \lbrace \left(-\muc \right)^m F(t_{i}-m\dtc) \rbrace\right]^2 , \\
  &= \sum_{i}^{N_D -1}\left[ \lbrace F(t_{i+1}) - F(t_{i}) \rbrace - \muc \lbrace F(t_{i+1}-\dtc) - F(t_{i}-\dtc) \rbrace \right. \nonumber \\  
  &\qquad \left.+ \muc^2 \lbrace F(t_{i+1}-2\dtc) - F(t_{i}-2\dtc) \rbrace   - \muc^3 \lbrace F(t_{i+1}-3\dtc) - F(t_{i}-3\dtc)\rbrace +  \hdots  \right]^2 . \label{eq:epF}
 \end{align}
 
For convenience, let us define the difference in the successive observed joint flux as a separate time series, 
  \beq
 H(t_i)\equiv F(t_{i+1})-F(t_{i})\;. \label{eq:H}
 \eeq
 Using Eq. \eqref{eq:H} one can simplify Eq. \eqref{eq:epF}  to
  \begin{align}
 \epsilon (\dtc) &= \sum_{i}\left[ H(t_i) - \muc H(t_i-\dtc) + \muc^2 H(t_i-2\dtc) - \muc^3 H(t_i-3\dtc)    + \hdots \right]^2 =\sum_{i} \left[ \sum_{n}^{\infty}\left(-\muc \right)^n H(t_i-n\cdot \dtc)\right]^2.
 \end{align}
 Further expanding the squared term, we get
 \beq
 \begin{aligned}
  \epsilon (\dtc)&= \sum_{i} H(t_i)^2 -2 \muc \sum_{i}H(t_i)H(t_i-\dtc) +\muc^2 \left[ \sum_i H(t_i-\dtc)^2 + 2\sum_i H(t_i)H(t_i-2\dtc)\right] \\  
  &\qquad -2\muc^3 \left[\sum_i H(t_i) H(t_i-3\dtc)  + \sum_i H(t_i-\dtc)H(t_i-2\dtc)\right]+ \hdots \label{eq:ep_long2} \;.
  \end{aligned}
 \eeq

 Thus we can express $\ept$ as a power series in $\mu$,
 \beq
 \ept =\epsilon_0 + \muc \epsilon_1(\dtc) + \muc^2 \epsilon_2(\dtc) + \muc^3 \epsilon_3(\dtc) + \muc^4 \epsilon_4(\dtc) + \hdots \;, \label{eq:ep_series}
 \eeq
 where
 \begin{align}
  \epsilon_0 &= \sum_i H(t_i)^2\;, \label{eq:ep0} \\
  \epsilon_1 (\dtc) &= -2\sum_i H(t_i)H(t_i-\dtc)\;, \label{eq:ep1} \\
  \epsilon_2 (\dtc) &= \sum_i H(t_i-\dtc)^2 + 2\sum_i H(t_i)H(t_i-2\dtc)\;, \label{eq:ep2} \\
  \epsilon_3 (\dtc) &= -2\sum_i H(t_i) H(t_i-3\dtc)  - 2\sum_i H(t_i-\dtc)H(t_i-2\dtc) \;, \label{eq:ep3}
 \end{align}
 and so on.
Therefore, the fluctuation curve can be written in the following closed form,
 \beq
 \epsilon (\dtc)=\sum_{n=0}^{\infty} \left(-\muc \right)^n \sum_{m=0}^{\lfloor n/2 \rfloor} \left(2-\delta_{m,n-m}\right)\sum_{i}\left[H(t_i-m \dtc) H(t_i- (n-m) \dtc) \right]\;,\label{eq:epn}
 \eeq
where the {\em floor function} $\lfloor n/2 \rfloor$ returns the highest integer equal to or below $n/2$. Since $\mu < 1$, the contribution from the higher order terms gets suppressed rapidly. 

Let us now take a closer look at the difference series $H(t)$ (defined in Eq. \eqref{eq:H}), which can be recast as
  \begin{align}
  H(t_i) &= \left[ \f(t_{i+1}) - \f(t_{i}) \right] +\mut \left[ \f(t_{i+1} - \dtt) - \f(t_{i} -\dtt) \right] \;, \\
  &=\htI(t_i)+\mut ~ \htI(t_i-\dtt) \;, \label{eq:Hh}   
  \end{align}
where
\beq
\htI(t_i)\equiv \f(t_{i+1}) - \f(t_{i})
\eeq
can be regarded as another time series. For a uniformly sampled data, $H(t)$ and $\htI(t)$ are proportional to the derivative of $F(t)$ and $\f(t)$, respectively. In this article, we refer to $H(t)$ and $\htI(t)$ as the derivatives of $F(t)$ and $\f(t)$, respectively. However, for non-uniformly sampled data, the former two are just difference series and the mathematical arguments remain intact. From  Eq. \eqref{eq:Hh} note that $H(t)$ follows similar lensing equation as $F(t)$ in Eq. \eqref{eq:lensed_flux} but with $h(t)$ as the underlying time series. 

The QSO intrinsic light curve, $\f(t)$, follows a stochastic process that is `wide-sense stationary'; the mean and covariance properties remain constant over the time. It is easy to check that the time series $F(t)$, $h(t)$ and $H(t)$ are also wide-sense stationary. For a generic wide-sense stationary time series $X(t)$,
\begin{align}
\langle X(t)\rangle_t \equiv \frac{1}{N_D} \sum_i X(t_i) &\approx \langle X(t + \tc)\rangle_t \;, \label{eq:meanA} \\
\sigma^2_{X(t)}  \equiv \langle X(t)^2 \rangle_t - \langle X(t)\rangle_t^2 &\approx \langle X(t+\tc)^2 \rangle_t - \langle X(t+\tc)\rangle_t^2 \equiv \sigma^2_{X(t+\tc)} \label{eq:sigA}
\end{align}
and
\beq\label{eq:tshift}
\langle X(t + t_{c_1}) X(t + t_{c_2}) \rangle_t \approx \langle X(t) X(t + \left[t_{c_2} - t_{c_1}\right]) \rangle_t \approx \text{function of}~ |t_{c_2}-t_{c_1}| ~\text{only}\;,
\eeq
for any shift in time much smaller than the observed time range ($\tc, ~t_{c_1}, ~t_{c_2} \ll t_{\rm range}$).
Therefore, 
\beq
\langle H (t) \rangle_t =\langle F(t_{i+1}) \rangle_t - \langle F(t_{i}) \rangle_t \approx 0
\eeq
since we assume that the statistical properties of the joint light curve remain invariant under translations, i.e. $\langle F(t_{i+1}) \rangle_t \approx \langle F(t_{i}) \rangle_t$. 

Now that we assembled all the necessary tools, we proceed to explain the characteristics of the fluctuation curve that are described in Section \ref{sec:method} and illustrated in Fig.~\ref{fig:ep_demo}.  
\begin{itemize}
 \item First, notice that Eq. \eqref{eq:ep_long2} is invariant under $\dtc \to -\dtc$ as long as $|\dtc| \ll t_{\rm range}$, since $H(t)$ also follows Eqs. \eqref{eq:meanA} -- \eqref{eq:tshift}. This explains the symmetry in the $\epsilon(\dtc)$ curve with respect to $\dtc=0$. 
 
 \item The first term in Eq. \eqref{eq:ep_long2} or \eqref{eq:ep_series}, $\epsilon_0$, is constant (independent of $\dtc$) and hence can be ignored. The next leading order term $\epsilon_1(\dtc)$ is proportional to the correlation coefficient between $H(t)$ and $H(t-\dtc)$ defined as (also known as the auto-correlation function (ACF), see Eq. \eqref{eq:gen_acf})
 \beq \label{eq:Hcross}
 \acf (H;\dtc)=\frac{\sum_i \left[H(t)- \langle H(t)\rangle_t \right]\left[H(t-\dtc)- \langle H(t-\dtc)\rangle_t \right]}{\sqrt{\sum_i \left[H(t)- \langle H(t)\rangle_t \right]^2 \cdot \sum_i \left[H(t-\dtc)- \langle H(t-\dtc)\rangle_t \right]^2}} \approx \frac{1}{\epsilon_0}\sum_i H(t_i)H(t_i-\dtc)\;,
 \eeq
where one can identify that $\langle H(t)\rangle_t \approx \langle H(t-\dtc)\rangle_t \approx0$ and the denominator simply reduces to $\sum_i H(t_i)^2=\epsilon_0$. Therefore, Eq. \eqref{eq:ep_long2} can be approximated as 
\beq
\epsilon(\dtc)\approx \epo \left[1-2\muc \acf(H; \dtc)\right] + \mathcal{O}(\muc^2)\;. \label{eq:ep_approx}
\eeq

\item The auto-correlation function, $\acfh$, is maximised to unity at $\dtc=0$. This in turn minimizes the $\epsilon_1 (\dtc)$ term that dominates the $\dtc$ dependence of the fluctuation curve in Eq. \eqref{eq:ep_series} at the leading order in $\mu$ \footnote{This could also be obtained from the inequality,
\begin{equation}
     \epsilon_1(\dtc) = -2\sum_i H(t_i) H(t_i - \dtc) 
    \geq -\sum_i H(t_i)^2 - \sum_i H(t_i - \dtc)^2
    \approx - 2 \sum_i H(t_i)^2 = -2\epsilon_0 
\end{equation}
following the fact that $2xy \leq x^2 + y^2$. The equality holds when $H(t_i) = H(t_i - \dtc)$ for all $i$ requiring $\dtc=0$ unless $H(t)$ is a constant function. Therefore, $\epsilon_1(\dtc)$ is minimized at $\dtc=0$. Note that this argument does not require $\langle H(t)\rangle_t  \approx0$.
}. 
Therefore, one always observes a global minimum in the $\epsilon(\dtc)$ curve at $\dtc=0$ irrespective of the system being lensed or not: $\epsilon(0) \approx \epsilon_0 - 2\muc \epsilon_0 + \mathcal{O}(\muc^2)$. In fact, it is easy to calculate the exact value $\epsilon(0)=\epo/(1+\muc)^2$ as given in Appendix \ref{app:global_min}.

 \item  When the system is lensed, $\acfh$ is also maximised locally at $\dtc=\pm \dtt$ as the correlation function picks up excess power due to the matching of two sets of same intrinsic features that are separated by the true time delay in the joint light curve (see Appendix \ref{app:acf} for detailed derivations). Thus the fluctuation curve,  $\epsilon(\dtc)$, shows a pair of secondary minima at $\dtc=\pm \dtt$ following Eq. \eqref{eq:ep_approx}. When $\htI$ is pure white noise, $\acf(H;\dtc=\pm\dtt) \approx \mut/(1+\mut^2)$  as evident from Eq. \eqref{eq:acf_dtt_whitenoise} for long observation ranges. The exact height of the secondary minima in $\ept$ is calculated in Appendix \ref{app:sec_min} under this approximation. On the other hand, when the system is unlensed ($\mut=0$), we get no such secondary minima. Therefore, by detecting this pair of secondary minima, which are the dominant part of the {\em lensing features} in the fluctuation curve, one can identify the system as lensed. Simultaneously, the location of this minima pair allows us to estimate the time delay of the system.
 
 \item As the $\epsilon_i$ terms in Eq. \eqref{eq:ep_series}, which contain the $\dtc$ variable, are independent of the choice of $\muc$, the locations of the lensing features (local minima and maxima due to lensing) in the fluctuation curve $\epsilon(\dtc)$ is very much insensitive to $\muc$. (However, note that for a higher value of $\muc$ higher order terms in Eq. \eqref{eq:ep_series} also contribute which makes the curve more fluctuating as evident from the three panels of Fig.~\ref{fig:ep_demo}.)
 Therefore, we cannot determine the true magnification ratio in this approach, although the time delay can be recovered accurately \footnote{In an ideal condition where $\htI$ is uncorrelated in time, the observation range is long, data have negligible noise etc, one can in principle determine the magnification ratio from the height of the secondary minima at $\pm\dtt$ using Eq. \eqref{eq:ep_sec_2} (or from the height of the secondary peaks at $\acf(H;\pm \dtt)$). But this won't be reliable for the realistic QSO light curves with unknown correlation and especially in the presence of significant observational noise.}.  
\end{itemize}

\begin{figure}
 \centering
\includegraphics[width=\textwidth]{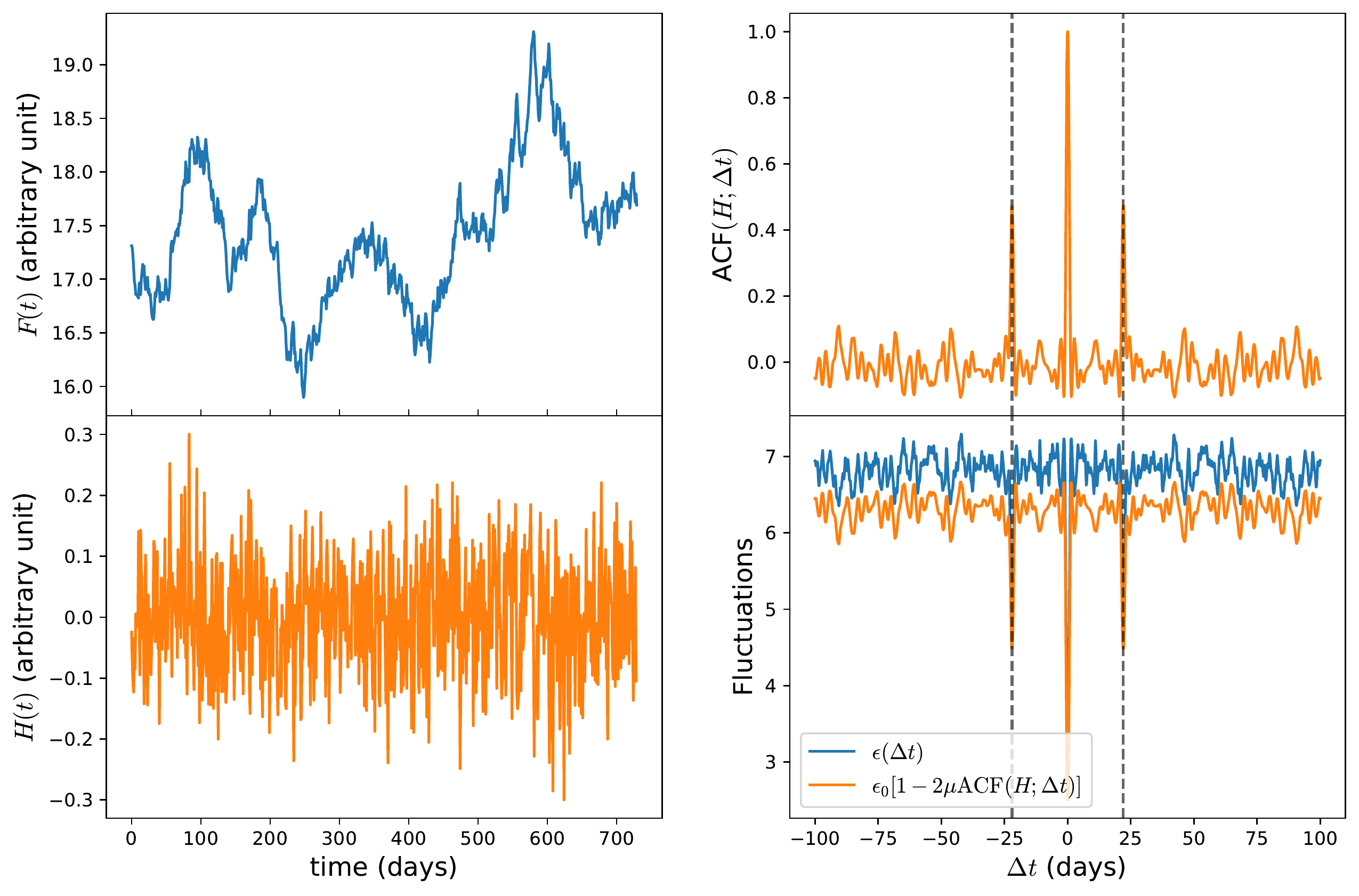}
\caption{The top-left panel shows the joint light curve (in perfect condition with negligible observation noise) of the double system considered in Fig.~\ref{fig:ep_demo}: $\dtt=22.0$ days and $\mut=0.86$. The difference series for this system, $H(t)$ as defined in Eq. \eqref{eq:H} (which is simply the derivative of $F(t)$ when it is sampled uniformly), is shown in the bottom-left panel. Its auto-correlation function, $\acfh$ shown in the  top-right panel, exhibits a prominent and sharp pair of secondary maxima at $\dtc=\pm\dtt$ through which one can easily detect the lensed system. Finally, the bottom-right panel compares the full fluctuation curve ($\epsilon(\dtc)$, same as in the left panel of Fig.~\ref{fig:ep_demo}) using the trial $\muc=0.3$ with the contribution from the $\acfh$ to it.
}
\label{fig:demo_acfh}
\end{figure}

Fig.~\ref{fig:demo_acfh} demonstrates the above arguments for the same example considered in Fig.~\ref{fig:ep_demo}. The top-left panel shows the joint light curve (in perfect conditions with negligible noise and one day of cadence) of the doubly lensed system simulated using the damped random walk (DRW) template with $\dtt=22.0$ days and $\mut=0.86$. The derivative of the joint light curve, $H(t)$, is shown in the bottom-left panel. The top-right panel displays the auto-correlation function of $H(t)$ where we can clearly find the pair of secondary maxima at $\dtc=\pm \dtt=\pm 22.0$ days. Finally, we compare the full fluctuation curve ($\epsilon(\dtc)$) with its main contributing term (linear in $\mu$) stemming from $\acfh$ for an arbitrary trial $\muc=0.3$. We notice that the features in the fluctuation curve are dominated by the $\acfh$, which is also responsible for the secondary minima in $\epc$ appearing at $\dtc=\pm \dtt$. The system is therefore identified as lensed. Note that the auto-correlation of the joint light curve itself, i.e. $\acff$, does not show the lensing peaks in this case; hence it is not reliable for lens detection as explained in Section \ref{sec:acff_acfh} in more detail.

\subsection{Contribution from the higher order terms}

For an arbitrarily long time series ($\tr \to \infty$), one can keep substituting
\beq\label{eq:long}
\sum_{i}^{N_D} H(t +\tc) H(t +\tc-n\dtc) \approx \acf (H;n\dtc)\cdot \epsilon_0 \;,
\eeq
for any $n$ ignoring the boundary effect. Here $\tc$ is any constant shift in time and $\epsilon_0$ is given by \eqref{eq:ep0}. Therefore, Eqs. \eqref{eq:ep1} -- \eqref{eq:epn}  can be simplified as 
\begin{align}
  \epsilon_1 (\dtc) &= -\epo \left[2\acf(H;\dtc)\right] \;, \label{eq:ep_long1} \\
  \epsilon_2 (\dtc) &=  \epo \left[2\acf(H;2\dtc) +1 \right] \;,\nonumber\\
  \epsilon_3 (\dtc) &= -\epo\left[2\acf(H;3\dtc)+2\acf(H;\dtc) \right] \;, \nonumber\\
  \epsilon_4 (\dtc) &=  \epo\left[2\acf(H;4\dtc) +2\acf(H;2\dtc) +1\right] \;, \nonumber\\
  &\hdots \nonumber \\
  \epsilon_n(\dtc) &=
  \begin{cases}
     \left(-1 \right)^n \epo\left[2\acf(H;n\dtc) +2\acf(H;(n-2)\dtc)+ \hdots +2\acf(H;\dtc)\right] & \text{if $n$ is odd}\;. \\
  \left(-1 \right)^n \epo\left[2\acf(H;n\dtc) +2\acf(H;(n-2)\dtc)+ \hdots +2\acf(H;2\dtc) + 1\right] & \text{if $n$ is even} \;.
\end{cases}
 \end{align}

Note that $\acf(H;n\dtc)$ exhibits a pair of peaks at $\dtc=\pm \dtt/n$.
Thus we get pairs of local minima (maxima) in the fluctuation curve at $\pm\dtt/n$ for each odd (even) $n$ from Eq. \eqref{eq:ep_series}. We refer to these features as the {\em lensing signal} in the fluctuation curve.
 Remarkably, from each odd term, we get a pair of minima at $\dtc=\pm \dtt$ which, despite being suppressed by a factor of $\muc^n$, enhances our target pair of secondary minima and increases its detectability. On the other hand, we find local maxima in the fluctuation curve at $\dtc=\pm \dtt/2, \dtt/4, \hdots$ from the even terms.  Therefore, as more and more terms contribute to the $\epsilon(\dtc)$ curve, we get slight enhancement in the lensing minima (as compared to the lensing peaks in $\acfh$) but at the expense of more extrema that increase the overall oscillations in the fluctuation curve. Note that for a finite time series, we still expect similar higher-order features in the fluctuation curve up to the order $n \ll t_{\rm range}/|\dtt|$.

 \subsection{Quad systems}
 \begin{figure}
 \centering
\includegraphics[width=0.485\linewidth]{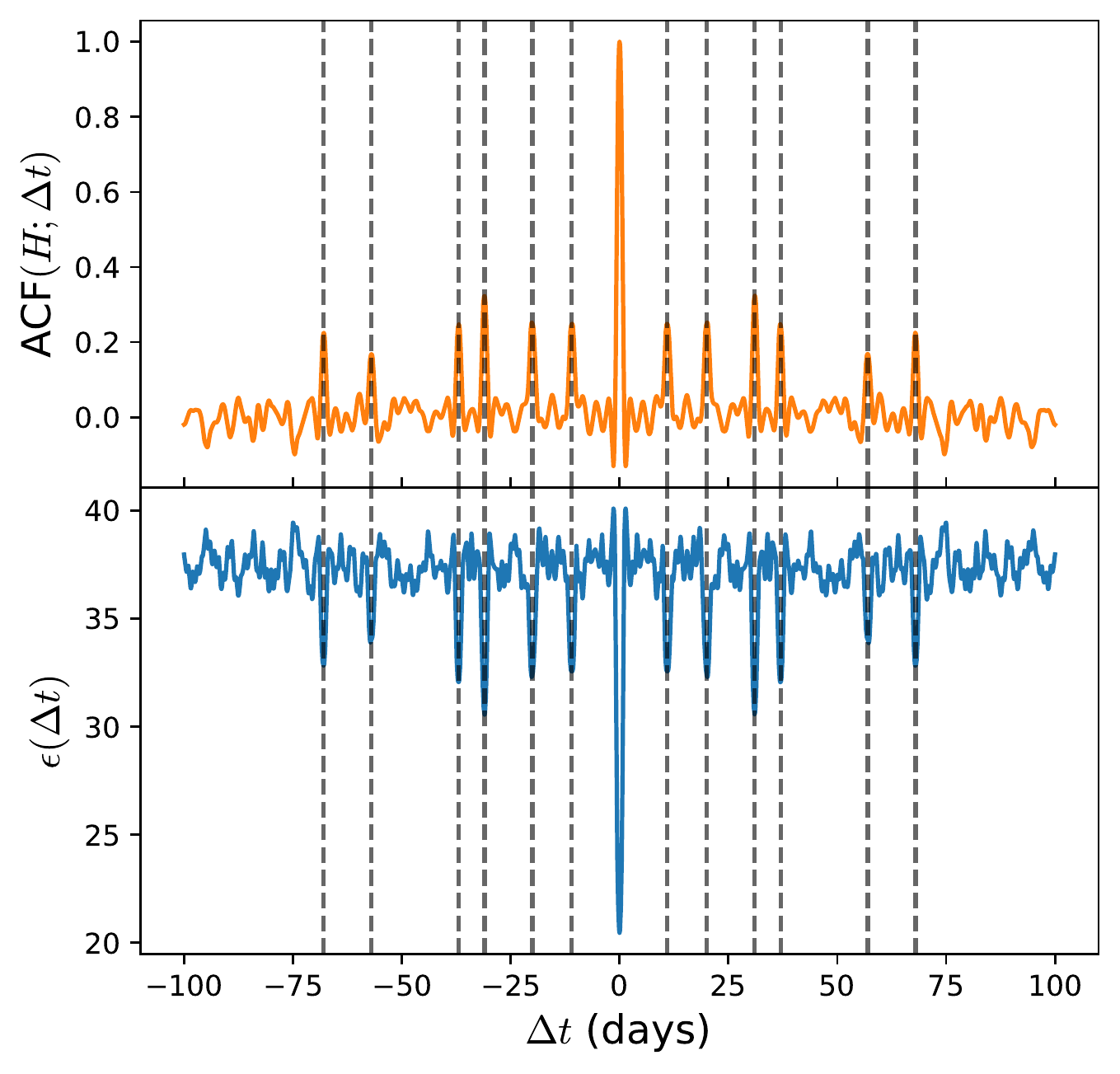}
\caption{The top panel shows the auto-correlation function of the derivative of the joint light curve ($H(t)$, defined in Eq. \eqref{eq:H}) for a quad system (4 images in reality).
The bottom panel shows the fluctuation curve for the same system, analysed assuming just two images. The six dashed vertical lines on either side of $\dtc=0$ in each panel represent the six relative time delays between the four actual images. }
\label{fig:quad}
\end{figure}

It is possible to generalise the image reconstruction, Eq. \eqref{eq:rec}, for more than two image systems. However, for a generic system with $N_I$ images, $\acfh$ exhibits $N_I (N_I-1)/2$ pairs of secondary maxima, one for each relative time delay $|T_i - T_j|$ ($i \neq j$). 
 Note that if the time delays between two pairs of images are too close to each other, or more explicitly, if the difference between two time delays is smaller than the observation cadence or the smoothing time scale (used to deal with noisy data), then the two corresponding peaks will merge into one in  $\acfh$.
For example, ACF$(H;\dtc)$ for a quad system (four images blended together in reality) can show up to six pairs of secondary maxima  at the time delays $\dtc= \pm \Delta t_{i,j\neq i}$ given that $\Delta t_{i,j\neq i}$ are well separated from each other.
As $\acfh$ dominates the lensing signal in the fluctuation curve, the latter also shows the same number of secondary minima pairs in this case even if one assumes only two images in the reconstruction analysis, i.e. following Eq. \eqref{eq:rec} with two images only. Thus, one can identify the quad systems using the method presented in paper 1 \citep{Bag2021qso} by detecting multiple (up to six) pairs of minima in the $\epsilon(\dtc)$ curve.

This has been illustrated in Fig.~\ref{fig:quad} where we consider an example of a quad unresolved lensed system simulated using DRW process with time delays: $\dtt=11.0, 31.0, 68.0$ days with respect to the first image (again in perfect condition). The top and bottom panels show the auto-correlation function for the derivative, ACF$(H;\dtc)$  and the fluctuation curve $\epsilon(\dtc)$ (using two image analysis), respectively. In both panels, we find six pairs of prominent extrema at $\dtc=\pm 11,~\pm 31,~\pm 68,~\pm 20,~\pm 57,~\pm 37$ days. Therefore, one can detect the quad systems using both approaches. In a blind analysis, if we detect one pair of secondary minima in the $\epsilon(\dtc)$ (or secondary maxima in $\acfh$) curve, we can identify the system as a doubly lensed QSO. On the other hand, if we find multiple such pairs (up to six), we can detect it as a quad system since 3-image systems do not exist in reality.

Thus, in this section, we explain the mathematical origin of the characteristics of the fluctuation curves which play the pivotal role in developing the method introduced in paper 1 \citep{Bag2021qso}. However, we are left with two additional questions: (i) why can one detect the lensed QSOs more reliably using $\acfh$ as compared to $\acff$, and (ii) what are the advantages of using the fluctuation curve over the simpler $\acfh$ in detecting lensed QSOs? These questions are addressed below in Sections \ref{sec:acff_acfh} and \ref{sec:fluc_vs_acfh}, respectively.

\section{$\acff$ vs $\acfh$}
\label{sec:acff_acfh}

 \begin{figure}
 \centering
 \subfigure[]{
\includegraphics[width=0.485\textwidth]{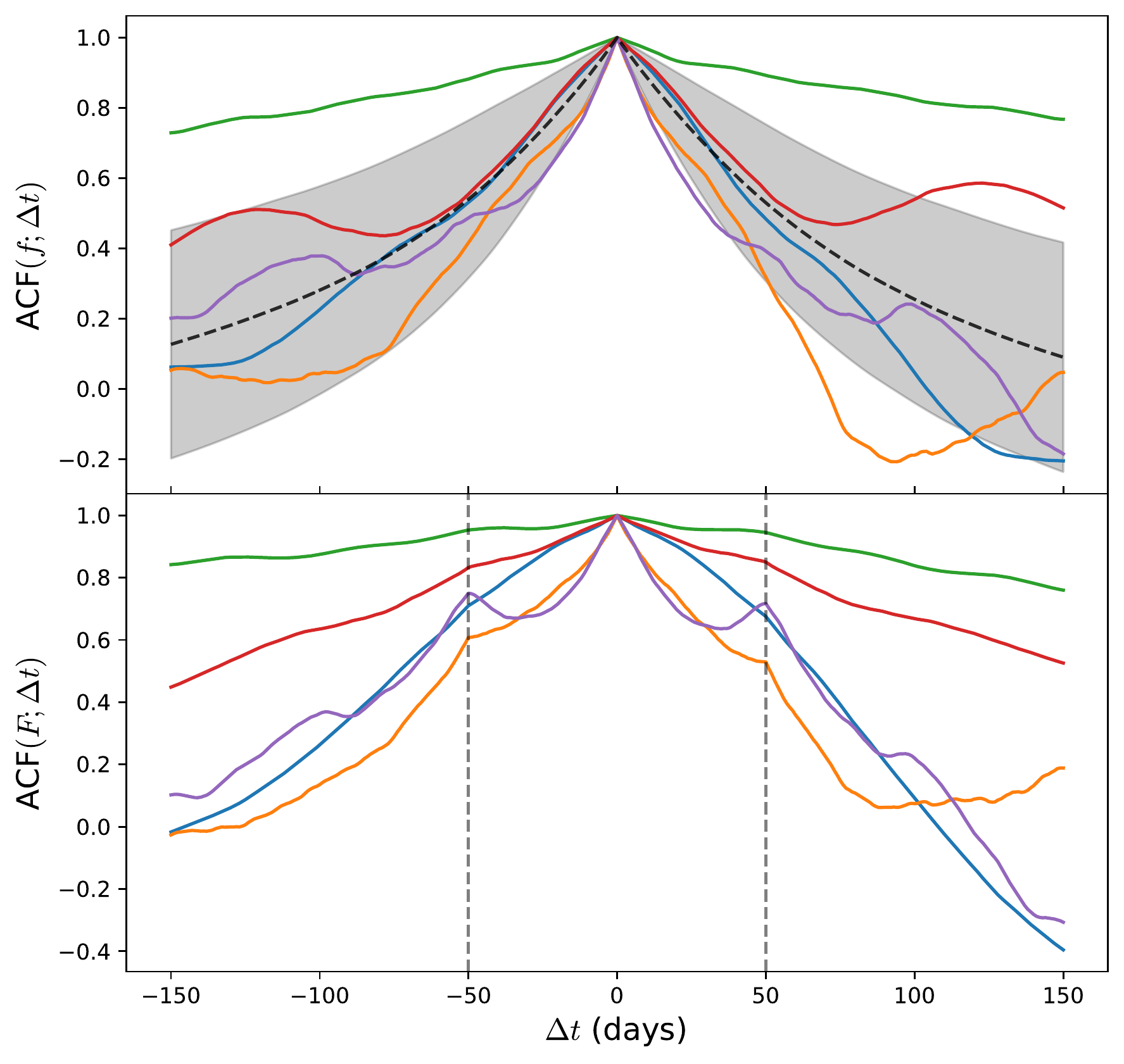}}
 \subfigure[]{
\includegraphics[width=0.485\textwidth]{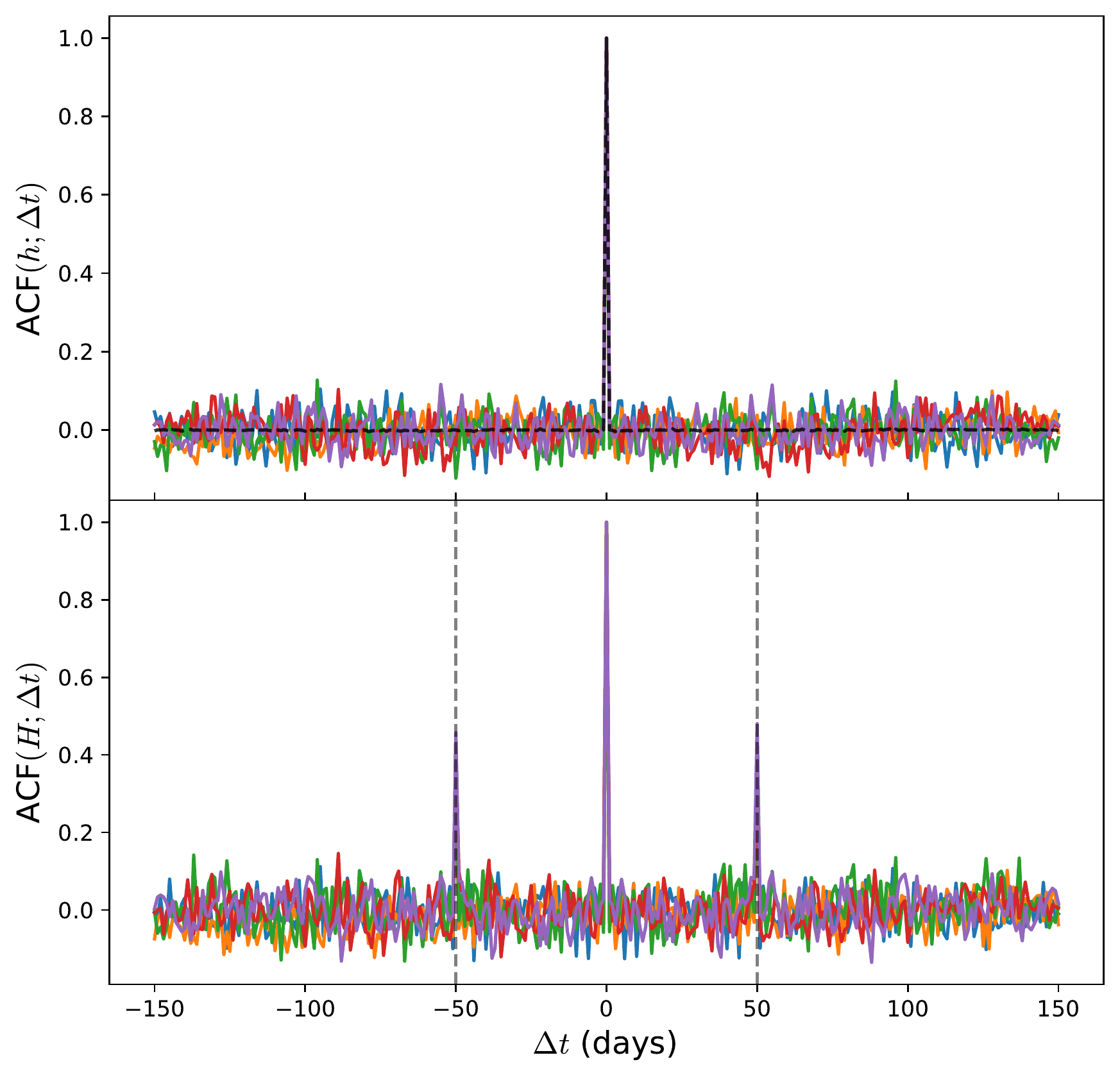}}
\caption{The top-left panel shows auto-correlation in the intrinsic QSO light curve, $\acffi$, for 5 out of 1000 realisations (randomly chosen) simulated using the DRW template. The expectation value (ensemble average) and $68\%$ percentile are marked by the dashed black curve and  the shaded region respectively. The top-right panel shows $\acfhi$ for the same five realisations along with the expected value (black dashed curve). 
The bottom two panels compare $\acff$ and $\acfh$ for the double systems constructed using these five realisations. The dashed vertical lines in the bottom two panels represent the true time delay: $\dtc=\pm 50$ days for these systems. Since $\acfhi$ is typically much narrower than $\acffi$, we expect to find the lensing peaks in $\acfh$ more reliably than in $\acff$, as evident from the comparison of the bottom two panels.
}
\label{fig:ACF_fFfH}
\end{figure}

One can in principle reliably detect the lensed systems using the auto-correlation function of the joint light curve, $\acff$, if the intrinsic light curve $f(t)$ is uncorrelated in time (white noise) as explained in Appendix \ref{app:acf}. However, the QSO light curves can have long time scale correlations that violate Eq. \eqref{eq:uncorr}. In this case, the existence of the lensing peaks in $\acff$ depends on the characteristics of the auto-correlation function of the intrinsic light curve, $\acffi$. 
Even when Eq. \eqref{eq:uncorr} is not strictly valid, we can expect excess power in $\acff$ at $\dtc=\pm\dtt$ from Eq. \eqref{eq:acfA11} if $\acf(f;\dtc)$ decays reasonably sharply (from its peak at $\dtc=0$), i.e. if $f(t)$ and $f(t \pm \dtc)$ are uncorrelated for $\abs{\dtc} \gtrsim \abs{\dtt}$. In other words, one can detect the lensing peaks in $\acff$ only if $\acffi$ is narrowly peaked around $\dtc=0$ even in the ideal condition with negligible observation noise. This is discussed in Appendix \ref{app:acff_peaks} in detail (see Eq. \eqref{eq:acf_condition} for the precise condition). Since $F(t)$ and its derivative $H(t)$ follow the same lensing equation (compare Eqs. \eqref{eq:lensed_flux} and \eqref{eq:Hh}), the above criterion is also applicable to $\acfhi$ in order to find the lensing peaks in $\acfh$.

 However, QSO flux variability typically shows temporal correlation till a few hundred to even thousand days \citep{Kelly2009, MacLeod2010}. The expectation value of the auto-correlation function of light curves generated using damped random walk (DRW) process \citep{Kelly2009,MacLeod2010,Zu13} decays exponentially,
\beq\label{eq:acf_drw}
\langle \acf(f;\dtc) \rangle_E = \exp(-\abs{\dtc}/\tau)\;,
\eeq
but not sharply since the decay time scale, $\tau$, is typically $\mathcal{O}(10^2 - 10^3)$ days. Here, $\langle \cdot \rangle_E$ denotes an ensemble average over all possible realizations of $f(t)$ with the same stochastic properties.

Therefore, it can be difficult to detect the secondary peaks in $\acff$, and one can also have a significant number of false positive detections \citep{Geiger1996,Shu2020}. This has been demonstrated in Fig.~\ref{fig:ACF_fFfH} using numerical simulations. Here we simulate $1000$ realisations of the intrinsic QSO light curves ($f(t)$) using the DRW template. We fix the correlation time scale to $\tau =10^{2.5} \approx 316$ days which is consistent with the findings of \citet{Kelly2009,MacLeod2010} (see also \citet{Dobler15}) throughout the paper for our illustration purpose. Then we construct the joint light curves ($F(t)$) separately for each realisation following Eq. \eqref{eq:lensed_flux} with $\mut=0.86$ and $\dtt=50$ days, kept same across the realisations. For simplicity, we consider the perfect condition with marginal noise in the data. The solid curves in the top-left panel show $\acf(f;\dtc)$ for five randomly selected samples, whereas the dashed back curve and the shaded region represent the ensemble average of $\acf(f;\dtc)$ and the $68\%$ quantile, respectively \footnote{Eq. \eqref{eq:acf_drw} is true only if the observation range is much larger than correlation scale, i.e. $t_{\rm range} \gg \tau$ ($\tau$ has been set to $10^{2.5}$ days for these simulations) so that Ergodicity is observed. So the black dashed curve in the top-left panel of Fig.~\ref{fig:ACF_fFfH} coincides with Eq. \eqref{eq:acf_drw} for much longer observation range.}. It is evident that there exists significant correlation in $f(t)$ till a few hundred days as the  $\acf(f;\dtc)$ curves decay slowly with $|\dtc|$. Also, notice that the $68\%$ quantile region of $\acf(f;\dtc)$ expands with $\abs{\dtc}$, thus some realisations of $\acffi$ can exhibit local maxima as in the case for the red and purple curves in the top-left panel. Thus, using $\acff$ one can get a substantial number of false lensed detections in true unlensed cases because of these maxima.

The $\acff$ for the same five realisations have been shown in the bottom-left panel, the dashed vertical lines represent the true time delay, $\dtc=\pm 50$ days, for these systems. Although for some realisations (purple curve) one can find excess power in $\acff$ at $\dtc=\pm \dtt$, for others (the green, red and blue curves) this is not true.

On the other hand, since DRW behaves like a random walk at small time scales ($t \ll \tau$), its derivative $h(t)$ behaves like white noise obeying Eq. \eqref{eq:uncorr} at this limit. This is evident from  the top-right panel of Fig.~\ref{fig:ACF_fFfH} where we show $\acf(h;\dtc)$ for the same five realisations with different colours. For all realisations we find that $\acf(h;\dtc)$ becomes very small for $\dtc\neq0$. In fact the ensemble average, shown by the dashed black curve, follows $\langle \acf(h;\dtc) \rangle \approx \delta_{\dtc,0}$. Hence, we get a prominent pair of secondary maxima in $\acfh$ at $\dtc=\pm \dtt$ for all the realisations as evident from the bottom-right panel.

Lastly, 
let us consider time series with different correlation time scales $\tau$ in Eq. \eqref{eq:acf_drw}, even if they do not describe the QSO light curves accurately. A smaller (larger) $\tau$ leads to a narrower (broader) peak in $\langle \acf(f;\dtc) \rangle$, which in turn increases (decreases) the probability of $\acff$ showing the lensing peaks, according to Appendix \ref{app:acff_peaks}. However, in both limits, $\acfhi$ has a sharp peak at $\dtc=0$ so that one can always find the lensing peaks in $\acfh$.     

This exercise using the DRW template thus demonstrates that $\acfh$ outperforms $\acff$ in terms of detectability of the lensing peaks. However, this conclusion is not restricted to DRW and unbound Random walk templates. It stands valid for any `red-type' power spectrum, as argued below in Section \ref{sec:psd} and in Appendix \ref{sec:psd_acfh_vs_acff} more explicitly.

\subsection{Connection to the power spectrum}\label{sec:psd}
Let us define the power spectrum, $P_f(\omega)$, of a time series as the two-point correlation function in the Fourier space;
\begin{equation}
    \langle \tilde{f}(\omega)  \tilde{f}(\omega')^* \rangle_E =P_f(\omega) \delta(\omega - \omega')  \;,
\end{equation}
where $ \tilde{f}(\omega)$ is the Fourier transform of the time series $f(t)$. We assume that the Fourier modes are uncorrelated. To be precise, we assume that the QSO intrinsic light curve $f(t)$ is `wide-sense stationary' as its mean and covariance properties do not vary over time (see \eqref{eq:meanA} -- \eqref{eq:tshift}). Under these assumptions, the Wiener-Khinchin theorem states that the (expected) auto-correlation function of $f(t)$ is given by the Fourier transform of the power spectrum \citep{wiener1930,Khintchine1934,Einstein1914}. Therefore, one can determine the auto-correlation function of a time series by studying its power spectrum. This is especially useful since a derivative in the time domain corresponds to a multiplication by $i\omega$ in the Fourier domain, and hence $P_{h}(\omega) = \omega^2 P_f (\omega)$.

The simplest example is when the underlying signal consists purely of white noise: $f_\textrm{white}(t)$ with the power spectrum given by ${P_{f_\textrm{white}}(\omega) = \sigma_\textrm{white}^2}=\text{constant}$. Taking the Fourier transform yields the expected value of the auto-correlation function ACF$(f_\textrm{white}; \Delta t) \propto \delta(\Delta t)$. Since this is sharply peaked around $\Delta t = 0$, we can accurately retrieve the lensing peaks in the $\acff$ curve. Further details for this white noise scenario are discussed in Appendix \ref{app:acf}.

For the damped random walk (DRW) templates often used for describing the intrinsic QSO light curves, the power spectrum takes the form
\beq\label{eq:drw_psd}
{P_{f_\textrm{DRW}}(\omega) \propto \tau/(1+\tau^2 \omega^2)}\;,
\eeq
which is the Fourier transform of Eq. \eqref{eq:acf_drw} up to a normalisation factor. Note that power spectra like $P_{f_\textrm{DRW}}(\omega)$ which decay with $|\omega|$ give rise to red noise and hence can be classified as `red-type' power spectra.
The decay time scale $\tau$ parameterises how long it takes for the time series to forget the fluctuations that happened in the past. Since $P_{f_\mathrm{DRW}} (\omega) \approx$ constant for $\tau^2 \omega^2 \ll 1$, the DRW is mostly independent of its past and behaves like a white noise when $\tau$ is small.

To describe realistic quasar light curves, $\tau$ should typically be of the order $10^2$ -- $10^3$ days. The DRW then behaves more similarly to an (undamped) Gaussian random walk with a power spectrum $P_f (\omega) \sim \omega^{-2}$ up to a time scale smaller than $\tau$, or $\omega^2 \gg 1/\tau^2$. The Fourier transform of such a power spectrum has a broad peak at zero unlike a sharp delta-function-like peak of the white noise one. Therefore, $\acff$ is less reliable for finding lensed cases in this case.

Meanwhile, the derivative time series $h(t)$ has a much flatter power spectrum with $P_h (\omega) \propto \omega^2 / (1+\tau^2 \omega^2)$. This is nearly constant for large $\omega$; $h(t)$ behaves like white noise. Thus, $\acfhi$ tends to be much narrower and is expected to decay down quickly as compared to $\acffi$. This is consistent with our previous findings that $\acfh$ is more reliable than $\acff$ for detecting the lensed systems. Indeed, a larger $\tau$ makes the $P_f(\omega)$ steeper (or redder) in Eq. \eqref{eq:drw_psd} and leads to a flatter $\langle \acffi \rangle$ in Eq. \eqref{eq:acf_drw}. This in turn reduces the possibility of the lensing peaks appearing in $\acff$ and thus enables $\acfh$ to outperform $\acff$ even to a greater extent.
In fact, this is true for all red type power spectra ($P_f(\omega)$ decreases with increasing $|\omega|$) as illustrated in Appendix \ref{sec:psd_acfh_vs_acff}. Since the lensing features in the fluctuation curve $\epsilon(\dtc)$ predominantly arise from $\acfh$ we expect similar performance from both these approaches.

\section{Fluctuation in reconstruction vs auto-correlation of derivative}
\label{sec:fluc_vs_acfh}

 \begin{figure}
 \centering
\includegraphics[width=0.485\textwidth]{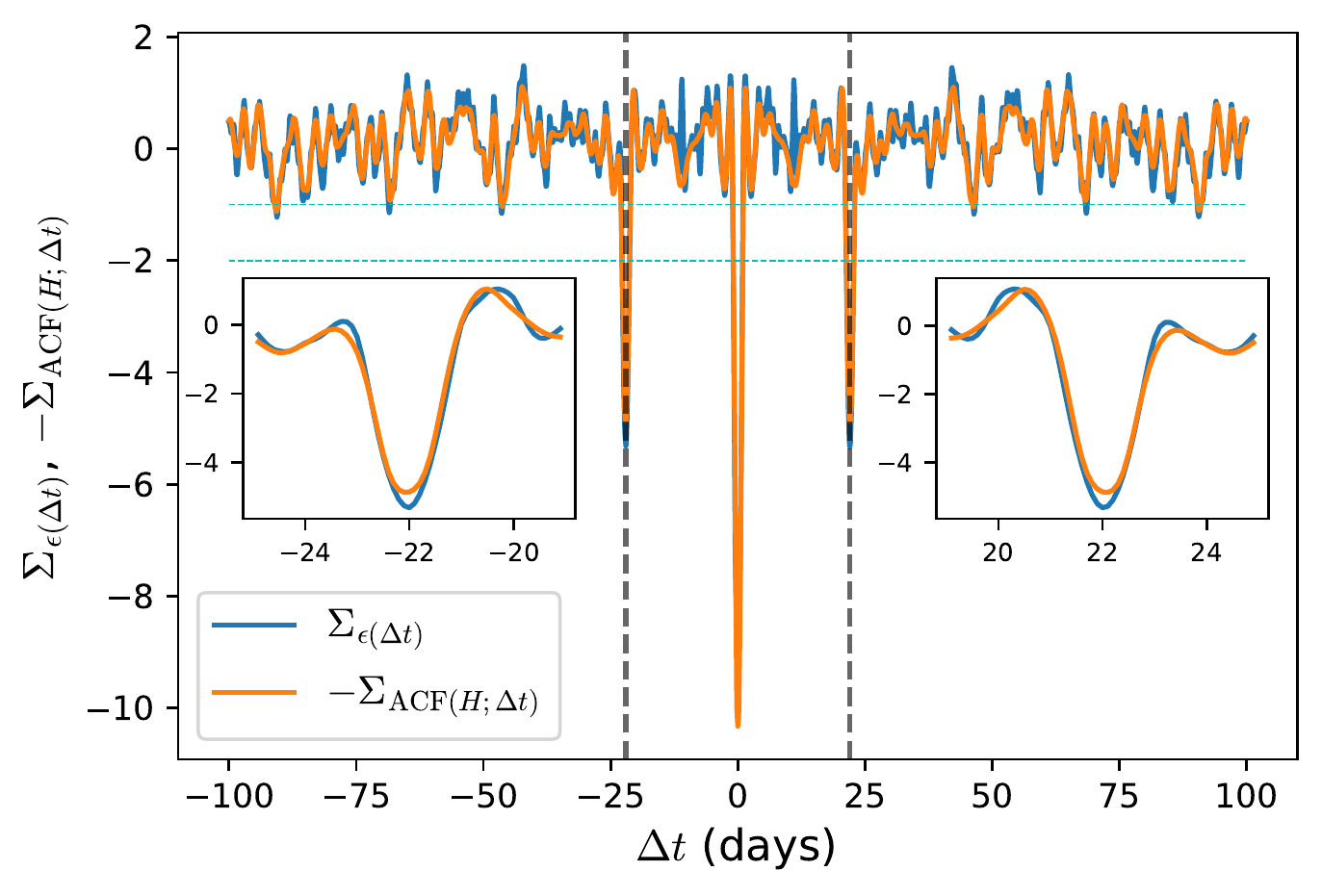}
\caption{The fluctuation curve and $\acfh$ are compared after being normalized using Eq. \eqref{eq:normalise} for the same system as in Fig.~\protect\ref{fig:ep_demo} (i.e. simulated with DRW template and with $\dtt=22.0$ days and $\mut=0.86$). We find secondary extrema of similar strengths in both curves as the $\acfh$ dominates the lensing signal in the fluctuation curve. However, the latter still has more lensing signal due to contributions from the higher order terms in Eq. \eqref{eq:ep_series} as illustrated by the two inset plots which zoom into the two lensing/secondary minima.}
\label{fig:Sigma_fluc_vs_ACFH}
\end{figure}

 Since the lensing signal in the fluctuation curve is dominated by the auto-correlation of the derivative of the joint light curve ($H(t)$ defined in Eq. \eqref{eq:H}), it is interesting to test if $\acfh$ can similarly be used to detect the lensed QSOs. Note that $\acfh$ would be less computationally expensive and has a better physical interpretation. 
 
 Let us begin by comparing the lensing signals (the prominence of the secondary maxima that can be used for lensed detection) in $\acfh$ and the fluctuation curve ($\epc$) for the example presented in Figs.~\ref{fig:ep_demo} (left panel) and \ref{fig:demo_acfh}. We normalize both curves using the generic transformation,
\beq\label{eq:normalise}
\Sigma_{X(\dtc)} \equiv \frac{X(\dtc) - \langle X(\dtc) \rangle_{\dtc} }{\sigma_{X(\dtc)}}\;,
\eeq
where $\langle X(\dtc) \rangle_{\dtc}$ and $\sigma_{X(\dtc)}$ are the average and standard deviation taken over $\dtc$ of a one-dimensional function $X(\dtc)$.
(Thus $\Sigma_{X(\dtc)}$ simply measures $X(\dtc)$ in the units of its standard deviation.) Fig.~\ref{fig:Sigma_fluc_vs_ACFH} compares $\Se$ (with trial $\muc=0.3$) and $-\SH$ for that example system (simulated using DRW template with $\dtt=22.0$ days, $\mut=0.86$ and marginal observation noise). As expected, both curves show prominent pair of secondary minima at $\dtc=\pm\dtt=\pm 22.0$ days (marked by dashed vertical lines) deeper than $\Sigma=-2$. However, we still find that the secondary minima in the $\Se$ are slightly deeper than that of the $-\SH$ curve. This indicates that $\epc$ contains a more enhanced lensing signal than $\acfh$ in this case.

  \begin{figure}
 \centering
\includegraphics[width=0.485\textwidth]{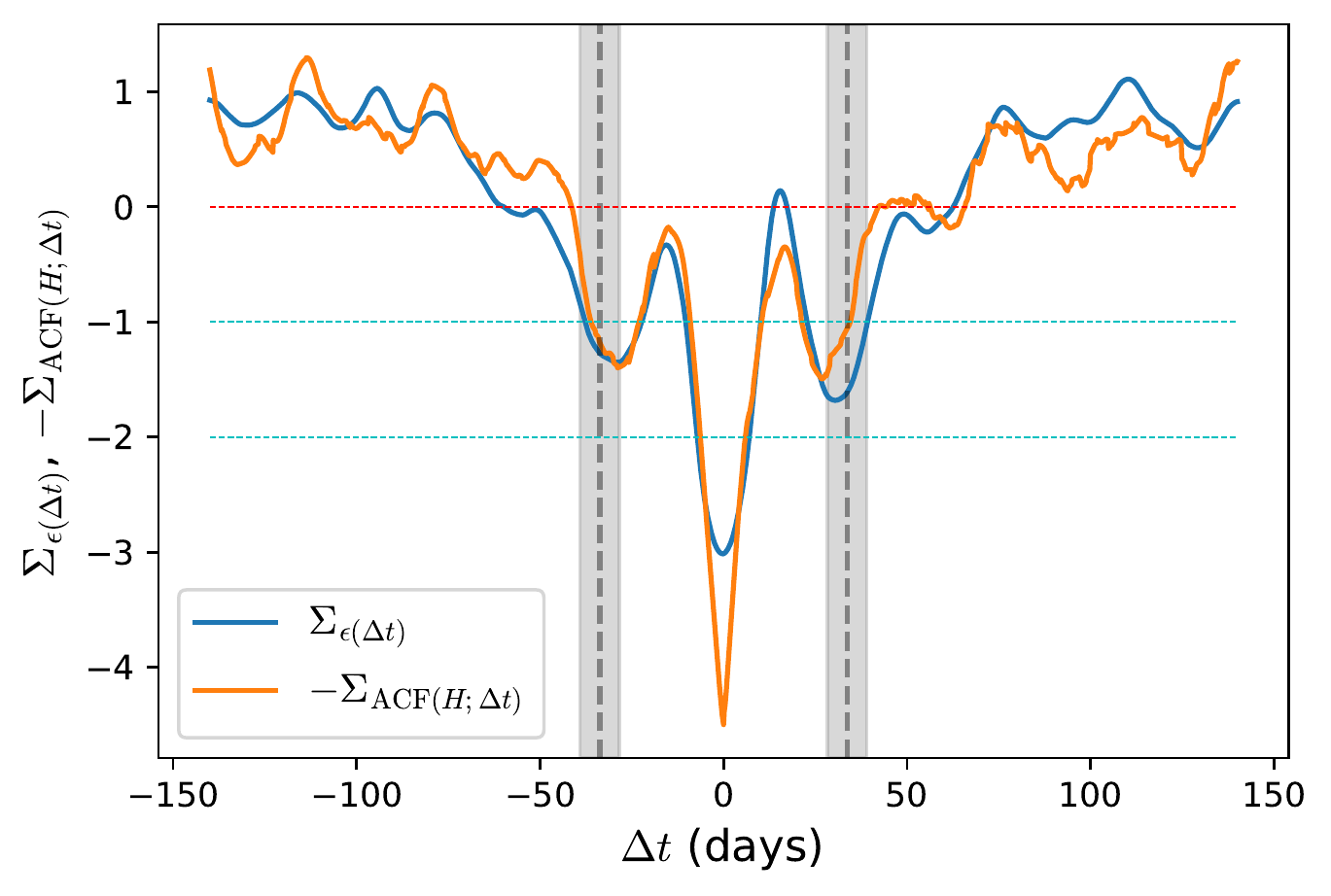}
\caption{The normalised fluctuation curve (blue curve, from \citet{Bag2021qso})  and $\acfh$ (shown by the orange curve) are compared for the  COSMOGRAIL system SDSS J1226-0006. 
}
\label{fig:J1226}
\end{figure}

 Next, we test the approach based on $\acfh$ on the same validation and blind sets used in \citet{Bag2021qso} (simulated using the DRW template). As usual, we first consider the perfect conditions where the observational noise is marginal compared to the time variation in the intrinsic QSO light curves. Using the same selection criteria, $\acfh$ can detect 19 out of 20 true lensed cases correctly. Among the 20 unlensed cases, however, it gives one false positive case while identifying the rest 19 unlensed systems correctly. In comparison, the minimization of the fluctuation (in the reconstructed image light curves) approach produces 1 false negative case but zero false positive cases for the same data sets. We also notice that the signal in $\SH$ is slightly diminished as compared to $\Se$. 
 
 When we consider uncertainty in the observed light curve data, the $\acfh$ approach suffers more from the added noise than the technique based on the fluctuation curve. We follow the same prescription given in \citet{Bag2021qso} for both methods to handle noisy data; we smooth the joint light curve for multiple smoothing scales and combine the fluctuation curves obtained from each smoothed light curve. For the same datasets with ZTF-like noise, $\acfh$ approach detects only 8 out of the 20 lensed systems correctly. However, for 2 other lensed cases, it detects the lensing nature based on peaks at completely wrong time delays (hence these two should be counted as false positives). Furthermore, it detects 15 out of 20 unlensed systems correctly but gives the rest 5 as false positives. Thus, in combination, it produces a precision of $8/15$ (slightly higher than $50\%$) and a recall of $8/20$. These numbers are significantly inferior to that of the fluctuation curve method which produces a precision of $12/13$ and a recall of $12/20$ on the same datasets \citep{Bag2021qso}. The recall and precision values for these two approaches have been summarized in Table \ref{tab:comp}. As before, we again notice that the lensing signal in $\SH$ is typically diminished as compared to $\Se$.
 
 \begin{table}
\centering
 \begin{tabular}{|c|c|c|c|c|} 
 \hline
  \multicolumn{1}{|c|}{\multirow{2}{*}{Datasets from }} & \multicolumn{2}{|c|}{Recall}& \multicolumn{2}{|c|}{Precision}\\\cline{2-5}
 & $\acfh$ & Fluctuation curve &  $\acfh$ & Fluctuation curve \\
\citep{Bag2021qso} &  & \citep{Bag2021qso}&  & \citep{Bag2021qso} \\
 \hline
With marginal noise & $ 95\%$ & $ 95\% $ & $ 95\%$ & $  100\%$ \\
 \hline
 With ZTF-like noise & $ 40\%$ & $ 60\% $ & $ 53.33\%$ & $ 92.31\%$ \\
 \hline

\end{tabular}
\caption{We compare the results from the two approaches -- $\acfh$ vs the full fluctuation curve -- in terms of recall (completeness) and precision (purity) for the validation and blind sets considered in \citet{Bag2021qso}. The results corresponding to marginal noise and ZTF-like noise in the data are presented in the top and bottom rows respectively.}
\label{tab:comp}
\end{table}
 
 We also test $\acfh$ on the COSMOGRAIL system SDSS J1226-0006 which was used in \citet{Bag2021qso} as an example. The time delay estimated using the resolved light curves by the COSMOGRAIL team is $33.7 \pm 2.7$ day for this system \citep{Millon:2020xab}.  Fig.~\ref{fig:J1226} compares the fluctuation curve (blue curve, same as in Fig. 21 of \citet{Bag2021qso}) and $\acfh$ (orange curve) after normalisation. Like the fluctuation curve, $-\SH$ shows a pair of prominent secondary minima symmetrically placed around $\dtc=0$. Specifically, the secondary minima in $-\SH$ occurs at $\dtc=-28.9,~26.9$ days with the depths $\Sigma=-1.40$ and $-1.50$ respectively leading to the final time delay estimation of $\dte=27.9$ days. In comparison, $\Se$ curve exhibits slightly overall deeper minima ($\Sigma=-1.35, -1.68$) at $\dtc=-28.7,~30.5$ days that give rise to $\dte=29.6$ days \citep{Bag2021qso} which is a better agreement with the COSMOGRAIL results.

 In conclusion, it is evident that the fluctuation curve approach as introduced by \citet{Bag2021qso} performs better than $\acfh$ although the former method is dominated by the latter. This is because the secondary minima (i.e. the lensing signal) are more prominent in the fluctuation curves, $\epsilon(\dtc)$, than the secondary maxima in the corresponding $\acfh$ curves. This is due to the fact that in $\epsilon(\dtc)$ curves, the secondary minima at $\dtc=\pm \dtt$ are further enhanced by all the odd terms in Eq. \eqref{eq:ep_series} as compared to $\acfh$ curves. Therefore, the minimization of fluctuation approach can be more useful for marginal detection of the unresolved lensed QSOs. However, $\acfh$ can be very useful and quick crosscheck as it is computationally inexpensive.

\section{Conclusion}
\label{sec:conclusion}

\citet{Bag2021qso} introduces a data-driven technique for detecting lensed QSOs and for measuring their time delays only using the unresolved joint light curve data by minimizing the fluctuations in the reconstructed image light curves. In this article, we provide the proof as to how this method works. We showed that the lensing signal in the simple fluctuation estimator given by Eq. \eqref{eq:ep_def} is dominated by the auto-correlation of the derivative (difference series in general for non-uniformly sampled data) of the joint light curve. This observation explains all the characteristics of the fluctuation curve that were used as the foundation of the technique proposed by \citet{Bag2021qso}. Above all, $\acfh$ is locally maximized at $\dtc=\pm \dtt$ and these lensing peaks manifest themselves as the secondary minima in the fluctuation curve, $\epsilon(\dtc)$, which have been used to detect the lensed cases in \citet{Bag2021qso}.
Other interesting results are summarized below.

\begin{itemize}
 \item We also showed that $\acfh$ is more reliable than the auto-correlation function of the joint light curve itself $\acff$, because the intrinsic flux variability of QSOs is correlated in the time domain, or in other words, the power spectra of the intrinsic quasar light curves are of red type. Nevertheless, even if $P_f(w)$ is flat or of blue type,  $\acfh$ can also find the lensed cases. Since the primary contribution to the lensing signal in $\epsilon(\dtc)$ comes from the $\acfh$, the minimization of the fluctuation approach would be similarly successful in these scenarios.

\item However, the approach based on the fluctuation curve provides better recall and precision over the $\acfh$ when one considers significant amount of noise in the joint light curve data. This is due to the higher-order terms contributing in Eq. \eqref{eq:ep_series} that further enhance the pair of secondary minima in the $\epsilon(\dtc)$ curve (as compared to the lensing peaks in $\acfh$).  
 
 \item For a generic lensed system having $N_I$ images, $\acfh$ displays $N_I(N_I-1)/2$ pairs of lensing local maxima. Likewise, the fluctuation minimization method can be used to detect multiple imaged lensed systems even if we assume two images in the reconstruction analysis a priori. For example, by detecting one pair of prominent secondary minima in $\epc$ one can identify a double system, whereas if there exist multiple such lensing minima pairs (up to six), the system must be a quad (having 4 images).
 
\end{itemize}

Although we choose time delays of the order of tens of days as examples in this work for demonstration purposes, one can in principle detect the lensing minima pair in the fluctuation curve for any arbitrarily small time delay as long as it is sufficiently larger than the observation cadence and the signal to noise ratio is sufficiently high. However, in reality, the cadence could vary from a few days to $\mathcal{O}(10)$ days depending upon the observation conditions or the observing strategy; this puts a limitation on the sensitivity of the method as the time delay needs to be larger than the cadence for the lensing minima pair to emerge in the fluctuation curve.

The fact that the primary contribution to the lensing signal in the fluctuation curves stems from $\acfh$ also informs us about some key benefits of the fluctuation minimization approach. 
The fluctuation minimization approach should be able to handle microlensing up to a certain limit as the auto-correlation can withstand moderate microlensing effects. We plan to comprehensively investigate the effect of microlensing on the performance of this method in the follow up work. Note that microlensing can significantly alter the time delay measurements from the resolved image light curves, up to a few days \citep{Tie2018,Kai2020}. It would be interesting to see how this affects the results of our method based on the unresolved fluxes. 

To discern another crucial advantage, recall that the selection criteria for lens detection using the fluctuation curve are so far based on only the pair of minima at $\dtc=\pm \dtt$.
The higher order terms further put predictable features in the fluctuation curve at certain values of $\dtc$, e.g. a pair of minima (maxima) at $\dtc=\pm \dtt/n$ for every odd (even) $n$. Although the higher order features are suppressed by $\mu^n$, the first few of these features can nevertheless be exploited to improve the selection criteria, potentially using deep learning.

Finally, we emphasise that the enhancement of fluctuations in the image light curves reconstructed using wrong time delays is a fundamental characteristic of the fluctuation curve approach. However, this article is restricted to the simple metric Eq. \eqref{eq:ep_def} for quantifying the fluctuations. The lensing signal in this estimator $\epsilon(\dtc)$ is found to be dominated simply by $\acfh$. Nevertheless, there might be a better metric for quantifying the fluctuations in the reconstructions that delivers better results in terms of recall and precision. This remains an open question and is worth investigating further.

 As stressed out in \citet{Bag2021qso}, to estimate the error in the time delay measurements in this non-parametric approach one needs to statistically analyse a large number of unresolved cases simulated in a variety of observational conditions; e.g. considering different flux variations, cadence distributions, noise levels, many microlensing realisations etc. This exercise forms the focus of the follow up work.

\section*{Acknowledgement}
We thank Tommaso Treu for his crucial inputs to this work. SB also thanks Eric V. Linder and Alex G. Kim for useful discussions. The Seondeok cluster at Korea Astronomy and Space Science Institute has been used to carry out a part of the analysis. A.S. would like to acknowledge the support by the National
Research Foundation of Korea NRF-2021M3F7A1082053 and
the support of the Korea Institute for Advanced Study (KIAS)
grant funded by the government of Korea. KL was supported by the National Natural Science Foundation of China (NSFC) under Grant Nos. 12222302, 11973034 and Wuhan University talent research start-up funds. A.S. and K.L. also acknowledge the support
and hospitality received from Beijing Normal University.

\appendix

\section{Auto-correlation function}
\label{app:acf}
The auto-correlation function of a generic time series $x(t)$ is defined as
\beq\label{eq:gen_acf}
\acfx=\frac{\sum_i^{N_D} \left[x(t_i)- \langle x(t)\rangle_t \right]\left[x(t_i-\dtc)- \langle x(t-\dtc)\rangle_t \right]}{\sqrt{\sum_i^{N_D} \left[x(t_i)- \langle x(t)\rangle_t \right]^2 \cdot \sum_i^{N_D} \left[x(t_i-\dtc)- \langle x(t-\dtc)\rangle_t \right]^2}}\;.
\eeq
For a wide-sense stationary (bounded and long) time series $\langle x(t)\rangle_t \approx \langle x(t-\dtc)\rangle_t$ which reduces the denominator to $N_D \sigma^2_x$. 
Let us define the joint time series $X(t)$ following the lensing equation (similar to $F(t)$ in Eq. \eqref{eq:lensed_flux} and $H(t)$ in Eq. \eqref{eq:Hh}) 
\beq\label{eq:X}
X(t)=x(t) + \mut x(t-\dtt)\;,
\eeq
where $\mut$ and $\dtt$ represent the (true) magnification ratio and relative time delay as in Eq. \eqref{eq:lensed_flux} and \eqref{eq:Hh}. Thus, $\lbrace x(t), X(t)\rbrace$ here are proxies for $\lbrace f(t), F(t)\rbrace$ or their derivatives $\lbrace h(t), H(t)\rbrace$. Using Eq. \eqref{eq:gen_acf} it is easy to find that
\beq\label{eq:acfA11}
\acfX=\frac{\acfx + \frac{\mut}{1+\mut^2}\left[ \acf(x;\dtc - \dtt) + \acf(x;\dtc + \dtt) \right]}{1+\frac{2\mut}{1+\mut^2}\acf(x;\dtt)}\;,
\eeq
where the denominator is just a normalisation constant (independent of $\dtc$).

If the intrinsic time series $x(t)$ is uncorrelated in time (white noise), sufficiently long and hence obeys
\beq\label{eq:uncorr}
\langle x(t)x(t-\dtc)\rangle_t=
\begin{cases}
   \langle x(t)^2\rangle_t &  \text{if } \dtc=0\;, \\
  \langle x(t)\rangle_t^2 & \text{if } \dtc\neq 0 \;,
\end{cases}
\text{~therefore,~} \acfx= 
\begin{cases}
    1 &  \text{if } \dtc=0\;, \\
   0 & \text{for all } \dtc\neq 0 \;,
\end{cases}
\eeq
the denominator of Eq. \eqref{eq:acfA11} reduces to unity (in the generic cases with $\dtt\neq0$) leading to
\beq\label{eq:acfA}
\acfX=\acf(x;\dtc) + \frac{\mut}{1+\mut^2}\left[ \acf(x;\dtc - \dtt) + \acf(x;\dtc + \dtt) \right]\;,
\eeq
and we can conclude the followings.
\begin{itemize}
 \item When $\dtc=0$, Eq. \eqref{eq:acfA} trivially reduces to unity as only the first term contributes. Naturally, the auto-correlation is always maximized at unity for no shift in time ($\dtc=0$).
 
  \item On the other hand, for $\dtc \neq 0$ in general, Eq. \eqref {eq:uncorr} ensures that all three terms in Eq. \eqref{eq:acfA} reduce to zero and $\acfX \approx 0$.
 
 \item Interestingly, $\acfh$ is locally maximized at $\dtc=\pm \dtt$. In the view of Eq. \eqref {eq:uncorr}, when $\dtc=\dtt$ or $-\dtt$ only the first or the second term in the square bracket of Eq. \eqref{eq:acfA} contributes, that leads to 
 \beq\label{eq:acf_dtt_whitenoise}
 \acf(X;\dtc=\pm \dtt)\approx \frac{\mut}{1+\mut^2} \leq \frac{1}{2}
 \eeq
 for both cases. In fact, the set of intrinsic features in the underlying time series $x(t)$ appears twice in $X(t)$, separated by $\dtt$ (and scaled by $\mut$, see  Eq. \eqref{eq:X}). Since $\acfX$ measures the correlation between two copies of $X(t)$ with one shifted by $\dtc$ in the time domain, it gains an excess power when $\dtc=\pm \dtt$ due to matching of the two sets of the same features ($\pm$ sign accounts for the shift in either direction). Although this argument applies to any generic $x(t)$ with a substantial amount of features, we emphasise that the above equation is valid strictly when the $x(t)$ is pure white noise following Eq. \eqref {eq:uncorr}. 
 
 \item For unlensed cases, when $\mut=0$ or $\dtt=0$, this pair of secondary maxima vanishes.

\end{itemize}
In summary, $\acfX$ shows a pair of secondary maxima at $\dtc =\pm \dtt$ (the lensing peaks) and remain zero at all other $\dtc \neq 0$ as long as the intrinsic time series, $x(t)$, is white noise (no temporal correlation) and satisfies Eq. \eqref {eq:uncorr}. 
Note that the whole analysis is applicable to both $F(t)$ and $H(t)$ with the underlying functions $f(t)$ and $h(t)$ respectively. 

\subsection{When to expect lensing peaks in $\acff$ even if $f(t)$ is not white noise?}
\label{app:acff_peaks}

Let us discuss the interesting case when Eq. \eqref {eq:uncorr} is not valid strictly. This is important for  quasars since the intrinsic light curves, $f(t)$, typically possess correlations up to a time scale of a few hundred days \citep{Kelly2009,MacLeod2010}. In such cases, the auto-correlation function $\acffi$ would show a broad peak at $\dtc=0$, unlike a delta function. $\acffi$ is still expected to be symmetric around $\dtc=0$ and monotonically decaying with $|\dtc|$ (the decay rate depends on the correlation time scale). For instance, see the top-left panel of Fig.~\ref{fig:ACF_fFfH} for typical examples of $\acffi$ where $f(t)$ is generated from damped random walk (DRW) with a correlation scale of $\sim 300$ days.

The lensing peaks in $\acff$ arise from the terms in the square bracket in the numerator of Eq. \eqref{eq:acfA11} (after replacing $\lbrace x(t), X(t)\rbrace$ by $\lbrace f(t), F(t)\rbrace$). Thus, as $|\dtc|$ approaches $|\dtt|$ from the below, the lensing peaks could emerge only if the change in the second term in the numerator of Eq. \eqref{eq:acfA11} dominates over the change in the first term. For all practical purposes (i.e. with the expected $\acffi$ being symmetric around $\dtc=0$ and monotonically decreasing with $|\dtc|$), this requirement boils down to the condition,
\beq \label{eq:acf_condition}
\left|\acf'(f;\dtc \sim \pm \dtt)\right| < \frac{\mut}{1+\mut^2} \left| \acf'(f;\dtc \sim 0) \right|\;,
\eeq
where $\acf'(f;\dtc)$ is the derivative of $\acf(f;\dtc)$ with respect to $\dtc$. The stronger the inequality is, the steeper (more prominent) the lensing maxima pair in $\acff$ would be. For our purpose, the above condition requires that $\acf(f;\dtc)$ must have sufficiently narrow peak at $\dtc=0$ so that it stabilizes at $\abs{\dtc} \gtrsim \abs{\dtt}$ by decaying down sufficiently fast. We again emphasise that all the arguments made in this section stand valid if we replace $F(t)$ and $f(t)$ by $H(t)$ and $\htI(t)$ respectively.
Note that here we assumed the best-case scenario when the noise is negligible;  the inclusion of observation noise brings in additional complexities. 

\section{Exact values of the global and secondary minima in the fluctuation curve}

\subsection{Global minima in the fluctuation curve}
\label{app:global_min}
For $\dtc=0$, Eq. \eqref{eq:eqF1} reduces to $\frI(t) = \frac{1}{1+\muc} F(t)$ which reduces Eq. \eqref{eq:ep_def} to  
\beq
\epsilon(\dtc=0) = \frac{1}{(1+\muc)^2} \sum_i^{N_D} \left[ F(t_{i+1}) - F(t_i) \right]^2 = \frac{\epo}{\left(1+\muc\right)^2}\;.
\eeq
Note that the above expression is valid even for the unlensed cases.

One can get the same expression from Eq. \ref{eq:ep_series} as follows,
\begin{align}
 \epsilon(\dtc=0) &=\epo\left[1-2\muc +3 \muc^2 - 4\muc^3 +5 \muc^4 -6\muc^5 + \hdots  \right]\;, \nonumber\\
 &=\epo \frac{\partial}{\partial \muc}\left[\muc -\muc^2 + \muc^3 -\muc^4 + \muc^5 -\muc^6 + + \hdots  \right]\;,\nonumber\\
 &=\epo \frac{\partial}{\partial \muc}\left[ \frac{\muc}{1+\muc}\right]=\frac{\epo}{\left(1+\muc\right)^2}\;.\label{eq:epdt0}
\end{align}
Here we identify that  for $\dtc=0$ all $\epsilon_n$ terms become proportional to $\epo$, and the sum converges as $\muc<1$.

\subsection{Secondary minima when $h(t)$ is white noise}
\label{app:sec_min}
If $\htI(t)$ is pure white noise (or in other words, if it follows Eq. \eqref{eq:uncorr}) we can analytically calculate the height of the secondary minima. Following Eq. \eqref{eq:ep_long1}, we get 

\begin{align}
\epsilon(\dtc=\pm \dtt)&=\epo \left[1-2\acf(H;\pm\dtt)\muc+\muc^2-2\acf(H;\pm\dtt)\muc^3+\muc^4-2\acf(H;\pm\dtt)\muc^5+\muc^6 - \hdots \right] \;, \\
&=\epo[\left(1+\muc^2+\muc^4+\muc^6 + \hdots \right) - 2\acf(H;\pm\dtt)\left( \muc+ \muc^3 + \muc^5 + \hdots \right)] \;,\\
&=\epo \left[\frac{1-2 \muc \acf(H;\pm\dtt)}{1-\muc^2}\right] \;, \label{eq:ep_sec_2}
\end{align}
where we use $\acf(H;n\cdot \dtt)\approx 0$ for $n \neq 0,~\pm1$. Note from Eq. \eqref{eq:acf_dtt_whitenoise} that $\acf(H;\pm\dtt) \approx \mut/(1+\mut^2)\leq 1/2$ in this limit. It is also easily seen that for unlensed cases, when $\acf(H;\dtc\neq 0)=0$, $\epsilon(\dtc)=\epo/(1-\muc^2)=\text{constant}$ for any $\dtc\neq 0$. 

By comparing with Eq. \eqref{eq:epdt0}, one can trivially show that 
\beq
\frac{\epsilon(\dtc=\pm \dtt)}{\epsilon(\dtc=0)}=\left(\frac{ 1+\muc}{ 1-\muc}\right) \left(1-2 \muc \acf(H;\pm\dtt) \right) \approx 1+2\muc\left(1-\acf(H;\pm\dtt)\right) + \mathcal{O}(\muc^2)\;,
\eeq
and owing to $\acf(H;\pm\dtt) \leq 1/2$ it is readily seen that $\epsilon(\dtc=0)<\epsilon(\dtc=\pm \dtt)$. Thus the central minimum at $\dtc=0$ will always be deeper than the pair of lensing minima.

\section{$\acfh$ is more reliable than $\acff$ for any red power spectrum}
\label{sec:psd_acfh_vs_acff}

\begin{figure}
 \centering
 \subfigure[]{
\includegraphics[width=0.485\textwidth]{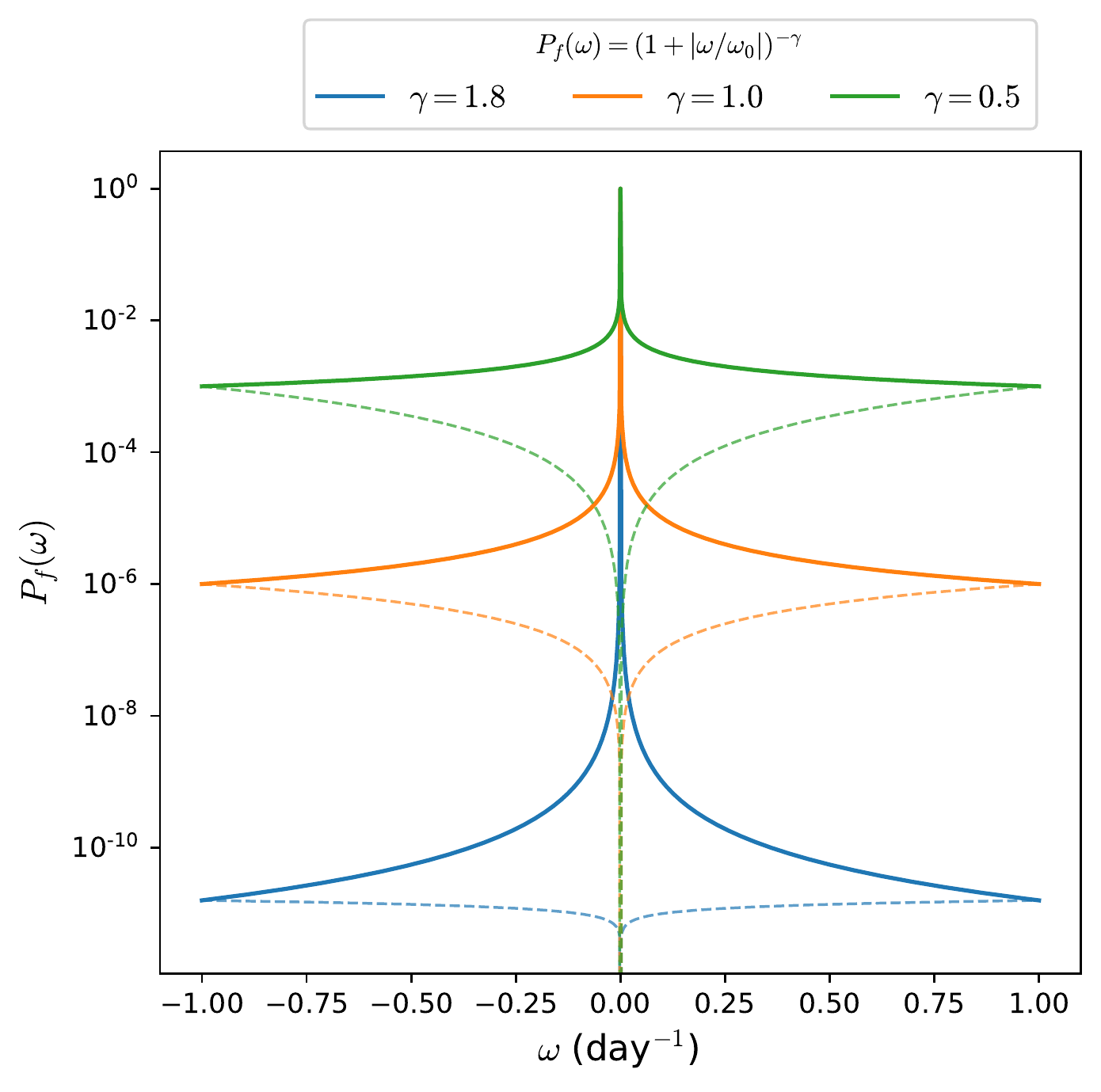}}
 \subfigure[]{
\includegraphics[width=0.485\textwidth]{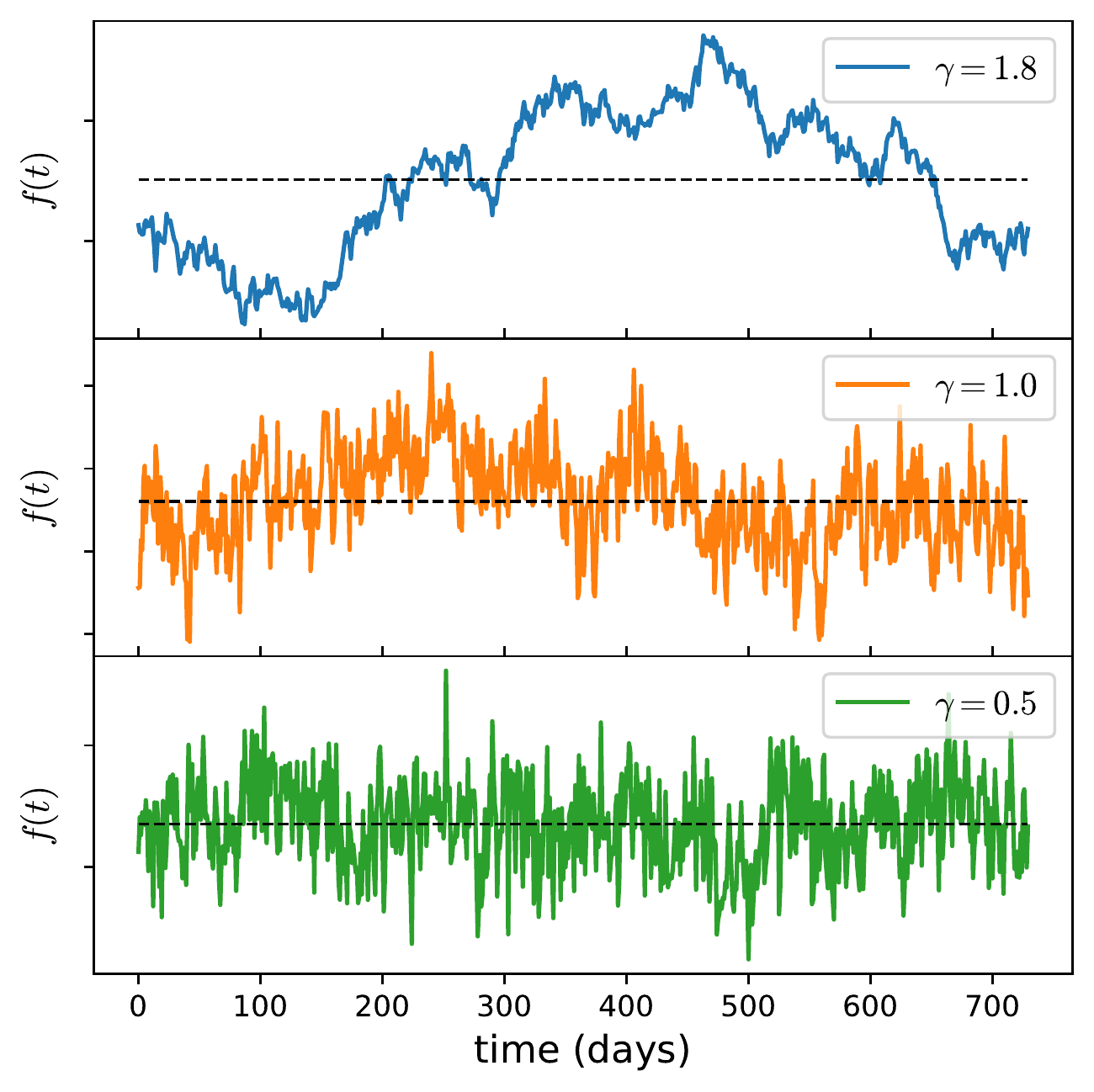}}
\caption{The left panel shows the power spectra ($P_f(\omega)$ in log-scale) given by Eq. \ref{eq:psd_example} for three values of $\gamma=1.8, 1.0$ and $1/2$ by the blue, orange and green solid curves respectively. Here we set $\omega_0=10^{-6}$ so that the power spectra behave like $|\omega|^{-\gamma}$ for the most part. The dashed curves represent the power spectra of the derivative: $P_h(\omega)=\omega^2 P_f(\omega) \approx |\omega|^{2-\gamma}$.
Each of the three panels on the right displays an example of the intrinsic flux $f(t)$ (in linear scale and arbitrary unit) generated from these power spectra. Note that observational noise is marginal here. The horizontal dashed line in each right panel represents the mean of the light curve.}
\label{fig:red_psd_examples}
\end{figure}
 \begin{figure}
 \centering
\includegraphics[width=\textwidth]{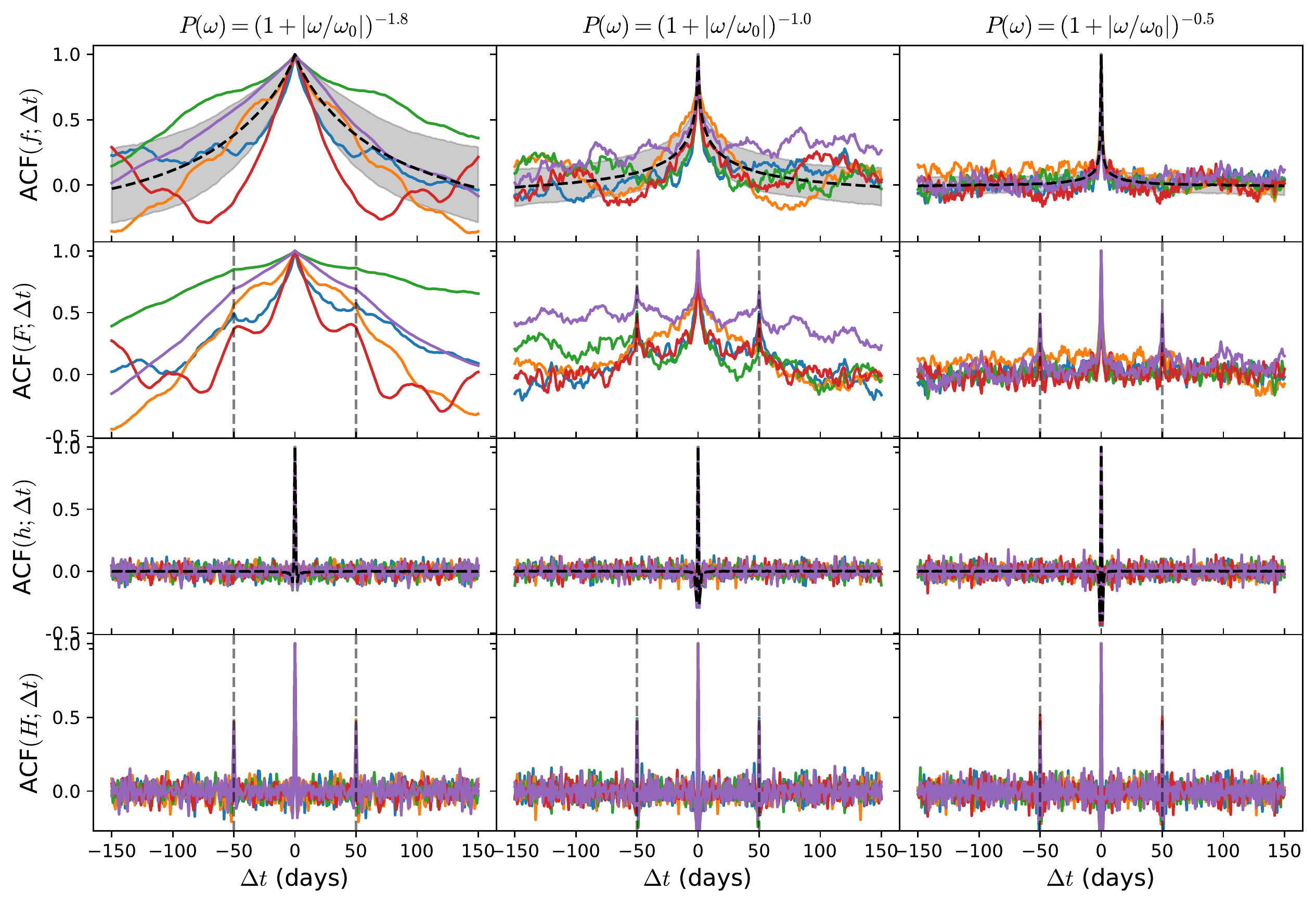}
\caption{Comparing $\acffi$, $\acff$, $\acfhi$ and $\acfh$ for five random realisations out of 1000 for three red type power-law power spectra: $P(\omega)=(1+\abs{\omega/\omega_0})^{-\gamma}$ with $\gamma= 1.8~\text{(left panels)}, 1~\text{(middle panels)}$ and $1/2~\text{(right panels)}$. Here, we set $\omega_0=10^{-6}$ day$^{-1}$  so that $P_f(\omega)$ behaves like a power-law for most part of it.}
\label{fig:psd_acf}
\end{figure}

We have seen in Section \ref{sec:acff_acfh} that the quasar intrinsic flux variability generated from a random damped walk (DRW) follows a red power spectrum given by Eq. \eqref{eq:drw_psd}. We found that $\acfh$ is more likely to exhibit the lensing peaks as compared to $\acff$ in this case due to the presence of temporal correlation in $f(t)$ up to a few hundred days realistically. Moreover, we discussed that a larger correlation scale ($\tau$) leads to a steeper (or redder) $P_f(\omega)$ in Eq. \eqref{eq:drw_psd} and a flatter (expected) $\acffi$ in Eq. \eqref{eq:acf_drw}, which enables $\acfh$ to outperform $\acff$ even more profoundly. In this appendix, we argue that this is not restricted to just the DRW template; for any time series with a red-type power spectrum, $\acfh$ is more reliable than $\acff$ for finding the lensed cases.

Recall that the expected auto-correlation function of a `wide-sense stationary' time series is given by the Fourier transform of the power spectrum. Since $\langle \acffi \rangle_E$ is symmetric around its peak at $\dtc=0$, the power spectrum should also be symmetric around $\omega=0$; $P_f(\omega)$ must be a function of $|\omega|$ only. Recall further from Eqs. \eqref{eq:acfA11} and \eqref{eq:acf_condition} that $\acff$ can show the lensing peaks only if $\acffi$ decays sharply from its peak at $\dtc=0$ and plateaus by $|\dtc| \gtrsim |\dtt|$. In other words, the narrower the (central) peak in $\acffi$, the better the chance of finding the lensing peaks in $\acff$. The same criterion applies to $\acfhi$ for the lensing peaks to appear in $\acfh$. However, for all practical purposes, a redder (or steeper) power spectrum, $P_f(\omega)$, leads to a flatter $\langle \acffi \rangle_E$ which diminishes the probability of finding the lensing peaks in $\acff$. A completely general proof of this statement is challenging, so here we argue using some concrete examples.

First, consider a Gaussian power spectrum;
\beq \label{eq:ftg}
P_f(\omega) \propto \exp(-a ~\omega^2/2)\;.
\eeq
Its Fourier transform is then also Gaussian with $\langle \acffi \rangle_E \propto \exp[{-\dtc^2/(2a)}]$. Thus, having a large $a$ in Eq. \eqref{eq:ftg} gives a narrow $P_f(\omega)$ but a broad $\langle \acffi \rangle_E$. This argument can be made more general; when we scale a function by $t \rightarrow at$, its Fourier transform gets inversely scaled,
\beq
\text{F.T}\left[g(at)\right]=\frac{1}{|a|}G(\omega/a)\;,
\eeq
where $G(\omega)$ is the Fourier transform of a generic function $g(t)$. Thus, if a scaling makes $P_f(\omega)$ steeper, $\langle \acffi \rangle_E$ would be flatter and vice-versa. On the other hand, the derivative series $h(t)$ has always a bluer power spectrum than that of $f(t)$ as $P_h(\omega)=\omega^2 P_f(\omega)$. Thus $\acfhi$ falls sharper from its peak at $\dtc=0$ as compared to $\acffi$ resulting in more prominent lensing peaks in $\acfh$.

Let us consider another example of red type power spectrum,
\beq\label{eq:psd_example}
P_f(\omega) \propto \left( 1 +|\omega/\omega_0|\right)^{-\gamma}\;,~\gamma>0\;,
\eeq
which behaves like a power-law for $\omega \gg \omega_0$ but avoids blowing up at $\omega \to 0$ by remaining stable at $|\omega| \ll \omega_0$. As the analytical solution for the Fourier transform of Eq. \eqref{eq:psd_example} does not exist for a generic $\gamma$, we carry out a numerical analysis. We set $\omega_0=10^{-6}$ day$^{-1}$ so that $P_f(\omega)$ behaves like $|\omega|^{-\gamma}$ in most part (except $\omega \approx 0$). In the left panel of Fig.~\ref{fig:red_psd_examples} we show three such power spectra with $\gamma=1.8, 1.0$ and $1/2$ by the solid blue, orange and green curves respectively. The power spectrum of the derivative, $P_h(\omega)=\omega^2 P_f(\omega) \approx  |\omega|^{2-\gamma}$, is shown by the dashed curve with respective colour. The three panels on the right show examples of the intrinsic light curves, $f(t)$, generated from the power spectra with the three values of $\gamma$ in Eq. \eqref{eq:psd_example}; the dashed horizontal line in each right-panel represents the mean of $f(t)$. It is evident that as $P_f(\omega)$ becomes redder with larger $\gamma$, $f(t)$ shows correlation till a longer time scale (i.e. two nearby points are more likely to be either above or below the mean unless not separated sufficiently in time).

Similar to the example illustrated in Fig.~\ref{fig:ACF_fFfH} for the DRW template, we now simulate 1000 realisations for the intrinsic flux variability separately for each of these three power spectra; the corresponding results are arranged in the three columns of Fig.~\ref{fig:psd_acf}. 
Then for each realisation, we construct a double lensed system using Eq. \eqref{eq:lensed_flux} with $\mut=0.86$ and $\dtt=50.0$ days, again we consider the perfect condition with marginal noise in the data for simplicity. The four rows in Fig.~\ref{fig:psd_acf} show $\acffi$, $\acff$, $\acfhi$, $\acfh$ respectively from the top. Five random realisations have been shown by the solid curves in each panel. The dashed black curve and the shaded region in the first and third row (from the top) panels represent the ensemble average and $68\%$ quantile around it. From the top-row panels, it is clearly evident that the redder the power spectrum (i.e. with larger $\gamma$) is, the slower the $\acffi$ decays from unity and the broader the $68\%$ quantile is. Therefore, a redder power spectrum in turn reduces the probability of detecting the lensing peaks in the $\acff$ as shown in the second row (from the top) panels. Also, statistically, we tend to get more false positive cases, e.g. the blue/red curve in the top-left panel.

On the other hand, for smaller $\gamma$ (flatter power spectrum), $\acffi$ tends to decay faster that increases the possibility of detecting lensed systems through $\acff$ (indeed, in the limit $\gamma \to 0$, $f(t)$ becomes white noise and $\acffi \propto \delta(0)$). Nevertheless, the peak prominence is still inferior to that of $\acfh$ (bottom panels). In contrast, for all the $\gamma$'s, $\acfhi$ decays sharply from its peak at $\dtc=0$ as evident from the dashed black curves in the third (from the top) row panels. \protect\footnote{When $\gamma=2$, $P_h(\omega) \approx |\omega|^{2-\gamma} =$ constant and $h(t)$ becomes white noise. Hence for $\gamma<2$, $h(t)$ possess (negative) correlation restricted to only the adjacent point that explains the local minima in $\acfhi$ just next to $\dtc=0$ on either side. But, this does not affect the detection efficiency of $\acfh$.} Thus, $\acfh$ shows the sharp lensing peaks at $\dtc=\pm \dtt$ (marked by the dashed vertical lines) for all the realisations and for all the power spectra considered. Therefore, Fig.~\ref{fig:psd_acf} (along with the Fig.~\ref{fig:ACF_fFfH} for DRW process) demonstrates that for any red type power spectrum, $\acfh$ is more reliable than $\acff$ for finding the lensed systems and the advantage of using $\acfh$ over $\acff$ increases for a redder power spectrum \protect\footnote{For completeness, let us discuss if the power spectrum is flat or blue ($\gamma>0$) although it does not describe QSO light curves well.
In these cases, $\acff$ shows sharp lensing peaks and so does $\acfh$ since the correlation in $h(t)$ is limited to the neighbouring point(s). Thus both perform well in these scenarios. This is also illustrated in Appendix C of \citet{Bag2021qso}}. Since the lensing signal in the $\epsilon(\dtc)$ curve is dominated by ACF$(H;\dtc)$, the fluctuation statistics introduced in \citet{Bag2021qso} 
is also very successful in all red power spectra scenarios.

\bibliographystyle{aasjournal}
\bibliography{lensed_quasars}
\end{document}